\documentclass[lettersize,journal]{IEEEtran}
\usepackage{amsmath,amsfonts}
\usepackage{algorithmic}
\usepackage{algorithm}
\usepackage{array}
\usepackage[caption=false,font=normalsize,labelfont=sf,textfont=sf]{subfig}
\usepackage{textcomp}
\usepackage{float}
\usepackage{stfloats}
\usepackage{url}
\usepackage{verbatim}
\usepackage{graphicx}
\usepackage{cite}

\AtBeginDocument{%
  \usepackage[subtle]{savetrees}
}

\usepackage{microtype}
\usepackage{color,soul}
\usepackage{booktabs} 
\usepackage{hyperref}
\usepackage{amssymb}
\usepackage{mathtools}
\usepackage{amsthm}
\usepackage[capitalize,noabbrev]{cleveref}
\usepackage[textsize=tiny]{todonotes}
\usepackage{graphicx}
\usepackage{multirow}
\usepackage{array}
\usepackage{placeins}
\usepackage{float}
\usepackage{siunitx}    
\usepackage{cellspace}
\usepackage{makecell}
\usepackage{threeparttable}
\usepackage{orcidlink}

\newcommand{\descr}[1]{\smallskip \noindent \textbf{#1}}

\begin{document}

\title{Bridging Local and Federated Data Normalization in Federated Learning: A Privacy-Preserving Approach}




\author{Melih~Coşğun$^{\ddagger}$,~Mert~Gençtürk$^{\dagger}$,~and~Sinem~Sav$^{\dagger\dagger}$\\[0.5em]
$^{\dagger}$Bilkent University, Ankara, Turkey\\
$^{\ddagger\ddagger}$Corresponding author: \textbf{sinem.sav@cs.bilkent.edu.tr}\\[0.5em]
Emails: melih.cosgun@bilkent.edu.tr, mert.gencturk@ug.bilkent.edu.tr,
sinem.sav@cs.bilkent.edu.tr
}


\markboth{November~2025}%
{Shell \MakeLowercase{\textit{et al.}}: A Sample Article Using IEEEtran.cls for IEEE Journals}


\maketitle

\begin{abstract}

Data normalization is a crucial preprocessing step for enhancing model performance and training stability. In federated learning (FL), where data remains distributed across multiple parties during collaborative model training, normalization presents unique challenges due to the decentralized and often heterogeneous nature of the data.
Traditional methods rely on either independent client-side processing, i.e., \textit{local normalization}, or normalizing the entire dataset before distributing it to parties, i.e., \textit{pooled normalization}. 
Local normalization can be problematic when data distributions across parties are non-IID, while the pooled normalization approach conflicts with the decentralized nature of FL. In this paper, we explore the adaptation of widely used normalization techniques to FL and define the term \textit{federated normalization}. 
Federated normalization simulates pooled normalization by enabling the collaborative exchange of normalization parameters among parties. 
Thus, it achieves performance on par with pooled normalization without compromising data locality. 
However, sharing normalization parameters such as the mean introduces potential privacy risks, which we further mitigate through a robust privacy-preserving solution. 
Our contributions include: (i) We systematically evaluate the impact of various federated and local normalization techniques in heterogeneous FL scenarios, (ii) We propose a novel homomorphically encrypted $k$-th ranked element (and median) calculation tailored for the federated setting, enabling secure and efficient federated normalization, (iii) We propose privacy-preserving implementations of widely used normalization techniques for FL, leveraging multiparty fully homomorphic encryption (MHE).

\end{abstract}


\section{Introduction}\label{sec:introduction}

\IEEEPARstart{M}{achine} learning (ML) has become an integral part of numerous applications, ranging from healthcare and finance to entertainment and smart devices. However, the increasing prevalence of sensitive user data in training ML models necessitates stringent privacy protections. Federated learning (FL)~\cite{Konency2016fed, federatedLearning1} has emerged as a promising paradigm for addressing this challenge by enabling decentralized model training. In FL, data remains on local devices, and only model updates are shared with a central server, reducing the risk of exposing raw data and thereby enhancing user privacy. Despite its advantages, FL remains vulnerable to several privacy attacks~\cite{Hitaj2017,Wang2019,Melis2019,9735364,10024757}, prompting the emergence of \textit{privacy-preserving federated learning (PPFL)} research.

A critical aspect of building effective ML models is data normalization, which involves scaling or transforming input features to improve convergence and performance. Normalization reduces the influence of feature-scale disparities, enabling models to learn more effectively. Widely used normalization techniques, such as MinMax scaling, z-score normalization, and robust scaling, have consistently proven their utility in centralized machine learning workflows. However, adapting these methods to FL presents unique challenges. Thus, especially \textit{PPFL} approaches that employ differential privacy~\cite{truex2020ldp,WU2022362,mcmahan2018LSTM,Wei2020,wu2019value}, homomorphic encryption~\cite{9935302,9833648,spindle,poseidon,rhode,Bozdemir,10654543}, or secure multiparty computation~\cite{zheng2019helen,9187932,falcon,9139658,CHEN2024120481,10646833,9900067} require privacy-preserving data normalization prior to training.

\begin{figure*}[t]
    \centering
    \includegraphics[width=0.85\textwidth]{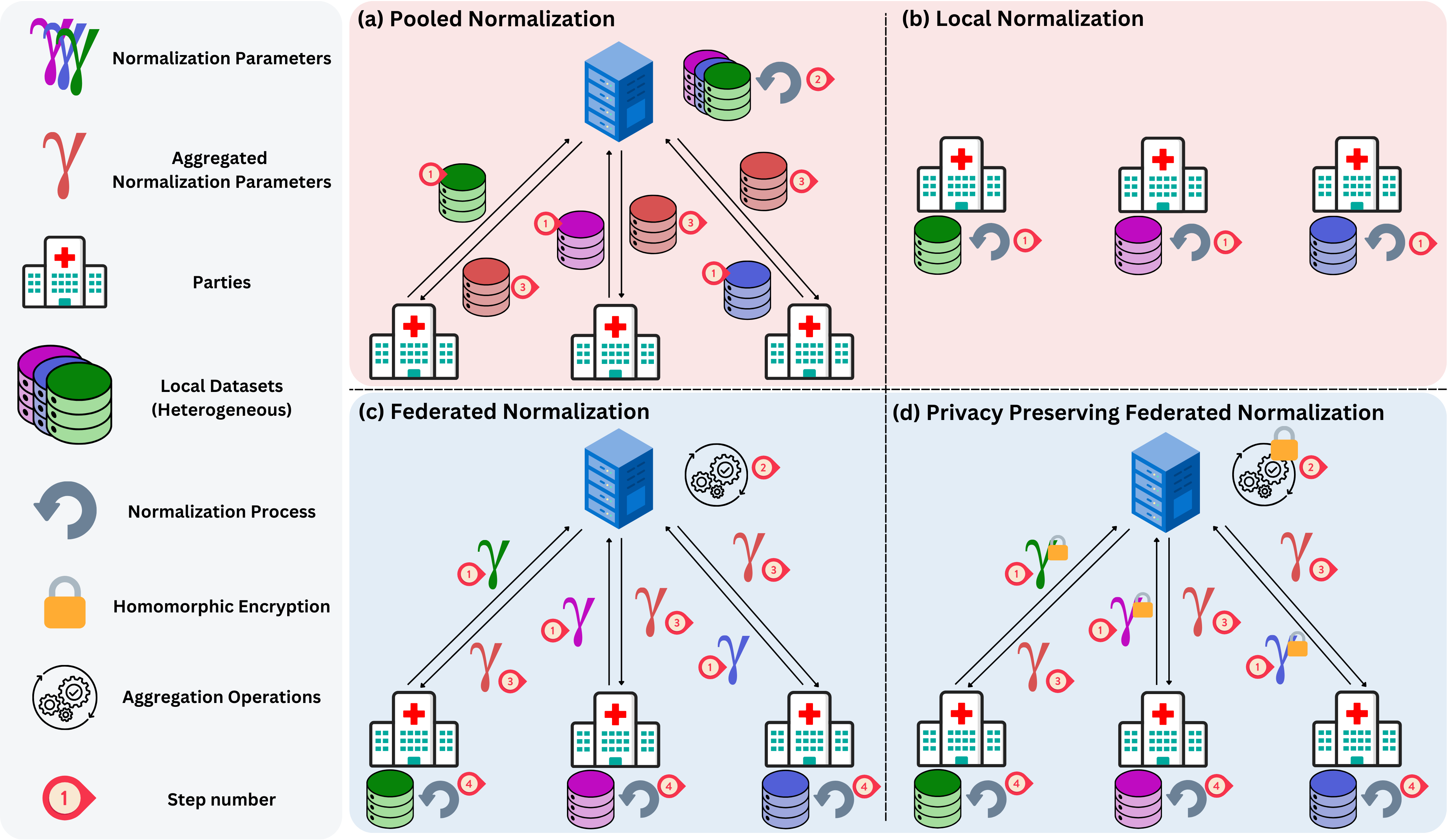} 
\caption{Overview of data normalization approaches in federated learning. The top (red) section shows traditional methods: (a) \textit{Pooled Normalization}, where raw datasets are centrally normalized, and (b) \textit{Local Normalization}, where each client normalizes data independently. The bottom (blue) section presents our federated methods: (c) \textit{Federated Normalization}, which aggregates normalization parameters without sharing raw data, and (d) \textit{Privacy-Preserving Federated (PPF) Normalization}, which enhances (c) using homomorphic encryption to securely aggregate parameters and minimize information leakage.}
    \label{fig:normalization_types}
\end{figure*}

In FL, the heterogeneity of client data and the lack of direct access to raw data make normalization a non-trivial task. Traditional approaches normalize datasets in two ways. The first approach, which we refer to as the \textit{pooled setting}, as also referred to in~\cite{securefedyj}, involves normalizing the entire dataset before distributing it to clients (see Figure~\ref{fig:normalization_types}-a). This method applies only to experimental settings, not real FL scenarios, as it assumes centralized data access, contradicting FL's decentralized nature. The second approach, which we refer to as the \textit{local setting}, involves each client normalizing their own data independently on their local device (see Figure~\ref{fig:normalization_types}-b). While this method aligns with FL principles and ensures data privacy, it encounters challenges in real-world scenarios where federated datasets are typically non-independent and identically distributed (non-IID). This non-IID nature leads to inconsistencies, reducing the effectiveness of local normalization compared to the pooled approach.

We define \textit{federated normalization} as normalizing federated data to achieve performance similar to pooled normalization by exchanging specific information among clients without sharing raw data. (see Figure~\ref{fig:normalization_types}-c). While this approach provides some privacy by keeping raw data local, the exchanged information can still potentially leak dataset insights, depending on the normalization technique. For example, federated MinMax scaling requires computing global feature-wise minima and maxima, which involves clients sharing their local extremes. However, this can risk revealing sensitive information, depending on the dataset and application context.

\descr{Problem Definition.} Existing literature lacks a thorough comparison between federated and local normalization approaches in FL. Key questions remain: What is the practical impact of federated versus local normalization on model performance, and how does data heterogeneity affect their effectiveness?
Finally, is it possible to implement federated normalization techniques in real-world settings while preserving client privacy?

To the best of our knowledge, the only prior work addressing a similar problem is~\cite{securefedyj}, which proposes a protocol for secure Yeo-Johnson (YJ) transformation in FL. While this work offers valuable insights, it is focused only on federated YJ Transformation and does not address the broader range of normalization techniques. Moreover, their evaluation is limited to regression tasks and does not fully address real-world heterogeneous FL scenarios. This leaves an opportunity for further investigation into the key questions we aim to address.
Similar to~\cite{securefedyj}, our work focuses on cross-silo FL, where a limited number of parties (typically tens to hundreds) collaboratively train a model. We use the terms 'clients' and 'parties' interchangeably throughout the paper.

\descr{Our contributions.} 
\begin{enumerate}
    \item For the first time, we present a comprehensive comparison of local and federated normalization techniques in non-IID FL scenarios. Our analysis spans diverse datasets and varying client counts, evaluating both local and federated settings. We demonstrate the critical role of federated normalization in enhancing FL performance.
    \item We introduce privacy-preserving implementations of various data normalization techniques, including z-score normalization, robust scaling, and MinMax scaling, tailored for FL. These implementations aim to reduce communication overhead while preserving client privacy. By leveraging multiparty fully homomorphic encryption (MHE), we ensure that client data remains confidential throughout the normalization process. 
    For an illustration of the core idea behind privacy-preserving federated (PPF) normalization, see Figure~\ref{fig:normalization_types}-d.
\item PPF robust scaling, one of our proposed privacy-preserving implementations, requires securely computing the \emph{$k$-th ranked element} among multiple parties. To address this, we propose a novel encrypted collaborative $k$-th ranked element calculation by extending the algorithm in~\cite{median} to support floating-point numbers. Leveraging MHE, our approach ensures data confidentiality with potential applications \textit{beyond normalization}.
\end{enumerate}

By addressing the intersection of data preprocessing and privacy-preserving machine learning, our work offers valuable insights and practical guidance to advance FL methodologies. Our implementation is available at \url{https://anonymous.4open.science/r/federated_normalization-9E66/README.md}.


\section{Related Work}
We summarize related work on local data normalization, activation layer normalization and PPF data normalization. We also discuss prior approaches for securely computing the $k$-th ranked element, relevant to our contributions in Section~\ref{sec:pp-normalization}.

\par\noindent\textbf{Local Data Normalization.}
Current research applies data normalization either in a pooled or local manner. Zhang et al.~\cite{healthcare_survey} note in their FL healthcare survey that normalization is typically performed before data distribution (pooled) or locally on each client. They further suggest that local normalization may propagate biases in feature values into the pre-processed data \cite{healthcare_survey}. Example FL works which rely on local or pooled data normalization include~\cite{local_example1, local_example2, local_example3, global_example1, global_example2}. To examine the impact of local normalization, we compare it against both no normalization and federated normalization scenarios.

\par\noindent\textbf{Activation Layer Normalization.} Several studies have investigated activation layer normalization in FL~\cite{batch1, wang2023batch, batch3}. Du et al.~\cite{batch1} introduce external covariate shift in FL and show that layer normalization outperforms batch normalization under non-IID conditions. Subsequent studies further explore batch normalization's limitations in FL and compare it with alternative normalization techniques~\cite{wang2023batch,batch3}. These normalization methods operate within neural network layers during training, in contrast to our focus on data normalization as a preprocessing step. Nonetheless, we include prominent activation normalization techniques—batch normalization, group normalization, instance normalization, and layer normalization—in our experiments. This allows for a comparative analysis of activation and data normalization methods across diverse FL scenarios, providing a comprehensive evaluation of their impact on model performance.

\par\noindent\textbf{Secure Computation of $k$-th Ranked Element.} The robust scaling normalization technique transforms data using the median, 25th percentile, and 75th percentile values. Thus, implementing robust scaling in a PPFL environment requires the secure and federated computation of $k$-th ranked elements (see Section~\ref{sec:pp-normalization}). 
Aggarwal et al.~\cite{median} compute the $k$-th ranked element in a multiparty setting via an iterative binary search, starting from a global range and querying data holders for counts above or below a candidate value. However, the method is limited to integers and lacks a secure implementation. Goelz et al.~\cite{median_implementation} address this by providing a secure multiparty implementation and benchmark, though it remains restricted to integer values and is not tailored for ML tasks. To support FL, we adapt this algorithm for floating-point inputs and incorporate MHE into its design. We introduce, for the first time, a federated and encrypted $k$-th ranked element calculation, designed to enable privacy-preserving federated normalization. To enhance security and minimize data leakage, we incorporate MHE into our design, which differs from previous work. Our implementation optimizes communication rounds by parallel processing of input features and is specifically designed for robust scaling in FL, ensuring both privacy and efficiency.

\par\noindent\textbf{Privacy-Preserving Federated (PPF) Data Normalization.}
The only related work by Marchand et al.~\cite{securefedyj} propose a secure implementation of the YJ transformation using secure multiparty computation within FL. Although YJ is an established normalization technique, its application is limited to specific tasks. In contrast, our work presents privacy-preserving implementations of additional widely used normalization methods, extending their applicability to a broader range of tasks and data types. In addition, the experiments in~\cite{securefedyj} are limited to regression tasks with numeric datasets due to the specific nature of the YJ transformation. Their evaluation primarily considers IID and quantity-imbalanced non-IID data distributions, which do not fully represent the broader spectrum of non-IID scenarios crucial for real-world FL applications. Furthermore, the datasets used in their experiments are relatively small (up to 1,797 samples), limiting their relevance for practical scenarios. As emphasized by~\cite{noniid}, incorporating diverse non-IID data distributions is essential to accurately reflect real-world FL challenges.

Our work addresses these limitations by investigating a wider range of non-IID data settings, incorporating both numeric and image datasets, and evaluating performance on regression and classification tasks. We systematically compare the proposed federated normalization techniques, including federated and local YJ transformations, across diverse non-IID scenarios using various neural network models. This comprehensive approach enhances the applicability of our findings, bridging important gaps and aligning more closely with practical FL deployments.

\section{Background}
\subsection{Normalization Techniques}  
Normalization is a key preprocessing step in machine learning that standardizes feature scales for effective training. For a dataset with $N$ features indexed by $j$, common normalization techniques are defined as:

\par\noindent\textbf{Z-Score Normalization:} This method scales features to have a mean of zero and a standard deviation of one. It is particularly effective when the data follows a Gaussian distribution. Formally, for a feature value $x_{j}$, the normalized value is given by: $x_{j}' = \frac{x_{j} - \mu_{j}}{\sigma_{j}}$, where $\mu_{j}$ and $\sigma_{j}$ are the mean and standard deviation of the $j$-th feature, respectively.

\par\noindent\textbf{MinMax Scaling:} This technique transforms data to lie within a fixed range, typically [0, 1], by applying the formula: $ x_{j}' = \frac{x_{j} - \min(x_{j})}{\max(x_{j}) - \min(x_{j})},$ where $min(x_{j})$ and $max(x_{j})$ represents the minimum and maximum values of the feature j. MinMax scaling is widely used when feature magnitudes vary significantly.

\par\noindent\textbf{Robust Scaling:} Designed to mitigate the impact of outliers, robust scaling scales features using the interquartile range (IQR) instead of the mean and standard deviation. The transformation is given by:  $ x_{j}' = \frac{x_{j} - \text{median}(x_{j})}{\text{IQR}(x_{j})}.$
\par\noindent\textbf{Yeo-Johnson Transformation:} The Yeo-Johnson transformation~\cite{yeojohnson}, an extension of the Box-Cox transformation~\cite{boxcox}, handles both positive and negative values. It stabilizes variance and helps data approximate a normal distribution. The transformation is defined as follows:
    \[
    y = 
    \begin{cases} 
      \frac{(x + 1)^{\lambda} - 1}{\lambda} & \text{if } x \geq 0, \\
      -\frac{(-x + 1)^{2 - \lambda} - 1}{2 - \lambda} & \text{if } x < 0,
    \end{cases}
    \]
    where $\lambda$ is a parameter that can be optimized to minimize the deviation from normality.

\par\noindent\textbf{Activation Layer Normalization:}
Unlike the previously mentioned methods, activation layer normalization techniques operate within neural network layers during training rather than in preprocessing. Prominent examples include layer normalization~\cite{layernormalization}, batch normalization~\cite{batchnormalization}, instance normalization~\cite{instancenormalization}, and group normalization~\cite{groupnormalization}. While sharing a common goal, these methods differ in their data processing approaches. They incorporate learnable parameters that integrate seamlessly into FL algorithms, similar to standard neural network components.

\subsection{Federated Learning with Non-IID Data}
Federated learning (FL) is a decentralized machine learning paradigm that enables model training across multiple clients without requiring data to leave their devices. Instead of sharing raw data, each client $i$ computes local models ${W}_{i}^{k}$ for $k$-th FL iteration and sends them to a central server. The central server then aggregates the local models to update the global model, e.g., ${W}^{k+1} \leftarrow \sum_{i=1}^{N} \frac{n_i}{n} {W}_{i}^{k}$ for FedAvg algorithm~\cite{federatedLearning1}. FL offers several advantages, including enhanced privacy, lower communication overhead, and the ability to utilize data that would otherwise be inaccessible due to privacy constraints.

In an \textit{IID (independent and identically distributed)} FL scenario, all clients’ data is assumed to be drawn from the same underlying distribution. This assumption simplifies the learning process, as local updates accurately reflect the global distribution, facilitating effective aggregation and faster convergence. However, real-world scenarios almost never satisfy the IID assumption: data is inherently non-IID~\cite{noniid, noniidIoT, 10833754, 9826416}, arising from diverse device types, geographic locations, user behaviors, or environmental conditions. This heterogeneity introduces several challenges that are critical to the success of FL.
Following Li et al.~\cite{noniid}, who proposed data partitioning to simulate common non-IID scenarios, we explore key non-IID cases to evaluate normalization techniques across diverse FL settings with $P$ parties:
\begin{itemize}
\item \textbf{Distribution-based Label Imbalance:} Simulates non-uniform label distributions across clients—for example, some may predominantly hold samples from certain classes, while others have a more uniform distribution. To simulate this, a Dirichlet distribution (\(\text{Dir}_P\)) is employed, in which each client \(i\) is assigned a proportion \(p_i\) of the samples for each label, with \(p_i \sim \text{Dir}_P(\beta_\ell)\). The concentration parameter \(\beta_\ell\) determines the level of imbalance: smaller values result in pronounced imbalance, while larger values yield more uniform distributions~\cite{noniid}.

\item \textbf{Noise-based Feature Imbalance:} The feature distributions differ across clients, often caused by variations such as environmental factors or sensor noise. To simulate this, the dataset is first randomly and equally partitioned among clients. Then, Gaussian noise \(n_i \sim \mathcal{N}(0, \beta_f \cdot i / P)\) with varying intensity levels is added to the features of each client \(C_i\)’s local dataset. The user-defined parameter \(\beta_f\) controls the level of noise added. This technique ensures that while the overall feature space remains consistent, the specific details vary significantly across clients~\cite{noniid}.

\item \textbf{Quantity Imbalance:} This case represents scenarios where FL clients possess datasets of varying sizes. This is simulated by allocating data samples to clients based on a Dirichlet distribution, where the concentration parameter dictates the extent of imbalance in the data quantities~\cite{noniid}.
\end{itemize}

These cases help us assess how non-IID data affects our proposed federated normalization and test the robustness of normalization methods in real-world scenarios.

\subsection{Multiparty Fully Homomorphic Encryption}\label{mhebackground} 

Fully homomorphic encryption (FHE)~\cite{fully_homomorphic_encryption} enables computations directly on encrypted data without decryption. It supports both addition and multiplication on ciphertexts, producing correct results upon decryption—as if the operations were performed on the original plaintexts.
One prominent FHE scheme, \textbf{CKKS (Cheon-Kim-Kim-Song)}~\cite{cheon2017homomorphic}, encodes plaintexts into polynomials packed into "slots", allowing parallel processing of multiple values. Designed for approximate arithmetic on complex numbers, it supports operations on floating-point arrays with minimal noise—ideal for machine learning tasks that tolerate slight approximation errors.

In this work, we use the MHE variant of CKKS~\cite{mouchet2021multiparty}, which distributes the encryption key to multiple parties. No single party can decrypt the data alone; decryption requires partial shares from all participants. This collaboration protects data unless all parties agree to decrypt, enhancing privacy and preventing unauthorized access. MHE is especially relevant to FL, enabling secure computation across distributed data while preserving data privacy. The core CKKS functionalities we rely on are summarized below:
\begin{enumerate}  
    \item \textbf{Key Generation }(\textsf{KeyGen($\lambda$)}): Each party $i$ generates a secret key $sk_i$ and participates in a secure key exchange to establish a collective public key $pk$ depending on the security parameter $\lambda$.  
    \item \textbf{Encryption} (\textsf{Encrypt($m,pk$)}): Message $m$ is encrypted using $pk$. We denote the encrypted message as $\langle m \rangle$.
    \item \textbf{Computations:} Homomorphic operations such as multiplication or addition are performed on the encrypted data. The resulting ciphertexts remain valid for decryption under the MHE scheme. However, repeated homomorphic operations can cause ciphertext noise to grow, potentially leading to decryption errors, which is eliminated with collective bootstrapping.    
    \item \textbf{Collective Bootstrapping }(\textsf{CBootstrap($\langle m \rangle,\{sk_i\}$)}): To reduce noise accumulation and enable deeper computations, the protocol uses collective bootstrapping~\cite{mouchet2021multiparty,Bossuat2021}, where all parties jointly refresh ciphertexts.
    \item \textbf{Collective Decryption }(\textsf{CDecrypt($ \langle m \rangle,\{sk_i\}$)}): Each party generates a partial decryption share using $sk_i$, which is then securely aggregated to reconstruct the computed result in plaintext.  
\end{enumerate} 

We omit the details of certain functionalities, such as relinearization and rescaling, provided by the MHE-CKKS scheme, which facilitate efficient management of ciphertext dimensions and scaling factors during computations.

\section{Evaluation of Normalization Techniques for Federated Learning}

To address the questions outlined in Section~\ref{sec:introduction}, we design a comprehensive experimental setup to evaluate the performance of local normalization, activation layer normalization, and our proposed federated normalization techniques. This section introduces federated normalization, outlines the experimental setup and its rationale, and concludes with an analysis and discussion of the results, highlighting key takeaways.

\subsection{Federated Normalization}

Before presenting the experimental evaluation, we begin by introducing the concept of federated normalization and outlining our proposed methods. In federated normalization, clients share certain information about their datasets, which is then aggregated to compute global normalization parameters (see Figure~\ref{fig:normalization_types}-c). These global parameters can be used to normalize data locally, ensuring that the data never leaves the client devices while enabling normalization as if the data were centrally aggregated.

Since the process is straightforward, we use MinMax as a representative example and present all algorithms, along with their PPF variants, in detail in Section~\ref{sec:pp-normalization}. In the MinMax scaling scenario, each client shares the minimum and maximum values of its local dataset for each feature. The central server aggregates these values to determine the global minimum and maximum. These global statistics are then used to normalize each client’s local data, resulting in outcomes equivalent to normalization performed on a centrally pooled dataset.

We propose and evaluate federated versions of MinMax scaling, robust scaling, and z-score normalization. Additionally, we include a federated version of the YJ transformation and activation layer normalizations in our experiments to provide a more comprehensive analysis.

\subsection{Experimental Setup}

We run experiments on five datasets across diverse non-IID scenarios. Key parameters include dataset choice, non-IID types and imbalance levels, number of clients, and normalization techniques. For all experiments, we use the FedAvg~\cite{federatedLearning1}, with one epoch of local training and evaluation during training is performed on a server-side validation set. Note that the validation dataset can also be distributed without affecting loss results. We use early stopping based on validation loss, adjusting the patience parameter. Each experiment is run three times with different random seeds, and we report the average. Experiments are implemented using the Flower FL library~\cite{flower} or our custom simulation, depending on memory constraints. All experiments are run on NVIDIA GTX 1080 Ti GPUs and AMD Opteron(TM) 6212 CPU with 256GB RAM.

\descr{Datasets.}
We conduct our experiments on two image and three tabular datasets. Four of these are used for classification tasks, while one is used for regression. We select these datasets to ensure diversity in size, features, tasks, and model architectures. In particular, we use CIFAR10~\cite{cifarPaper}, MNIST~\cite{MNIST}, Breast Cancer Wisconsin \cite{bcw}, Hepatitis C \cite{hepatisis} and Parkinson's Monitoring \cite{parkinsons}. We reserve 20\% of the datasets for the unseen test set and split the remaining data into 90\% for training and 10\% for validation. Supplementary Material~\ref{sec:supdetailedModel}, Table~\ref{tab:dataset_summary} summarizes the dataset specifications.

\descr{Model Architecture.}
We employ neural network-based architectures to incorporate activation layer normalization techniques in our experiments.  For image datasets, we adopt convolutional neural networks (CNNs), while for tabular datasets, we use fully connected neural networks integrating variable activation layer normalization. Based on the normalization type, we use layer, group, batch, or instance normalization for the layers. We omit activation layer normalization from our networks during data normalization experiments. Model architecture details are provided in Supplementary Material~\ref{sec:supdetailedModel}, Table~\ref{tab:model_architectures2}.

\descr{Non-IID settings.} Following the dataset imbalance methods in~\cite{noniid}, we apply six non-IID and one IID setting per dataset. Non-IID types include feature, label, and quantity imbalance, each with high and low imbalance configurations (see Table~\ref{tab:skew}). For image datasets, we use different feature imbalance settings than for tabular data, as strong feature imbalance on images degrades quality and leads to model divergence.

\begin{table}[t]
    \centering
    \begin{threeparttable}
    \renewcommand{\arraystretch}{1.2}
    \setlength{\tabcolsep}{6pt}
    
    \caption{Summary of imbalance parameters}
    \label{tab:skew}
    \vspace{1mm}
    \begin{tabular*}{\linewidth}{@{\extracolsep{\fill}}ccc}
        \toprule
        \textbf{Imbalance Type} & \textbf{Low Imbalance} & \textbf{High Imbalance} \\
        \midrule
        Quantity & $\beta = 0.5\ (0.7)^*$ & $\beta = 5$ \\
        Label    & $\beta = 0.5$         & $\beta = 5$ \\
        Feature  & $\sigma = 0.3\ (0.01)^*$ & $\sigma = 0.7\ (0.1)^*$ \\
        \bottomrule
    \end{tabular*}
    
    \begin{tablenotes}
        \footnotesize
        \item[$*$] Parameters in parentheses indicate values used for image data.
    \end{tablenotes}
    \end{threeparttable}
\vspace{-1em}    
\end{table}

\descr{Number of clients ($P$).}
We rely on a setting with 10, 20, and 50 clients for image datasets and 10, 20, 30 clients for tabular datasets-due to data size limitations.

\descr{Normalization Techniques.}
For all datasets, we evaluate both federated and local versions of z-score normalization, MinMax scaling, and robust scaling across all datasets. For tabular datasets, we also include local and federated YJ transformations. Additionally, we explore activation layer normalization methods, including layer, batch, group, and instance normalization. To efficiently simulate federated normalization, we normalize all training, validation, and test sets using parameters computed from the training set. For local normalization, each client computes normalization parameters from its local training data and applies them to its own training, validation, and test sets. After the normalization, validation and test sets are merged and moved to the server to speed up evaluation. 

\descr{Evaluation Metrics.}  For all datasets, we track the training loss curves for both decentralized and centralized setups, the validation loss curve, and the final test loss. For classification tasks, we additionally track training and validation accuracy curves, final test accuracy, F1 score, precision, and recall, using cross-entropy as the loss function. For regression tasks, performance is evaluated using mean absolute error (MAE) and $R^2$ on both validation and test sets, with MAE as the loss.

\subsection{Experimental Results}\label{sec:results}

\begin{figure*}[t]
\centering
\includegraphics[width=0.7\textwidth]{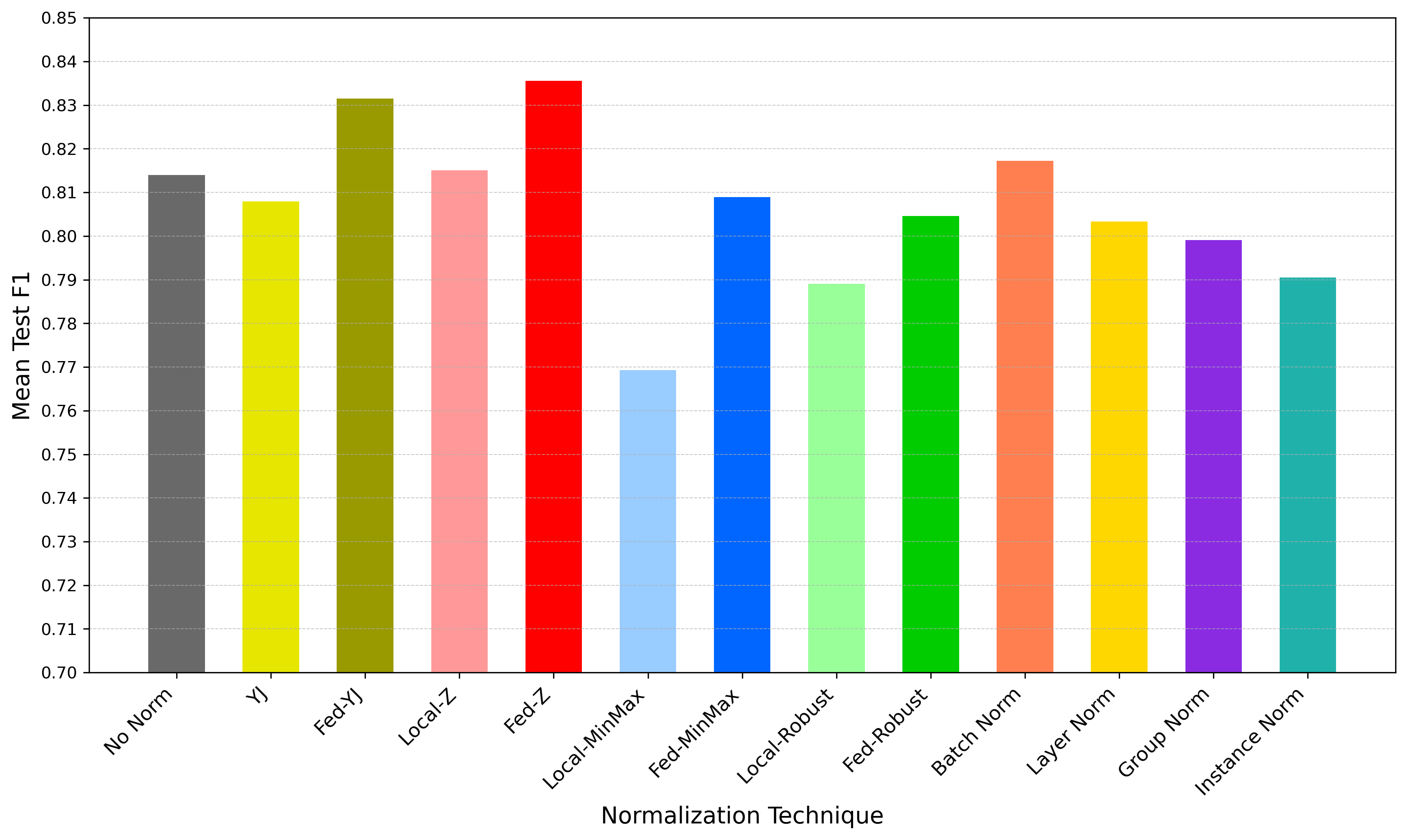}

\caption{Test F1 scores averaged across datasets, client numbers, and imbalance types for various normalization techniques. YJ transformations were applied only to tabular datasets. Only classification datasets are considered, as F1 scores represent classification performance. Labels prefixed with 'Fed' denote the federated versions of normalization methods, while 'No Norm' represents the baseline without normalization. }
\label{fig:avg_plot}
\vspace{-1em}
\end{figure*}

\begin{figure*}[t]
\centering
\includegraphics[width=0.9\textwidth]{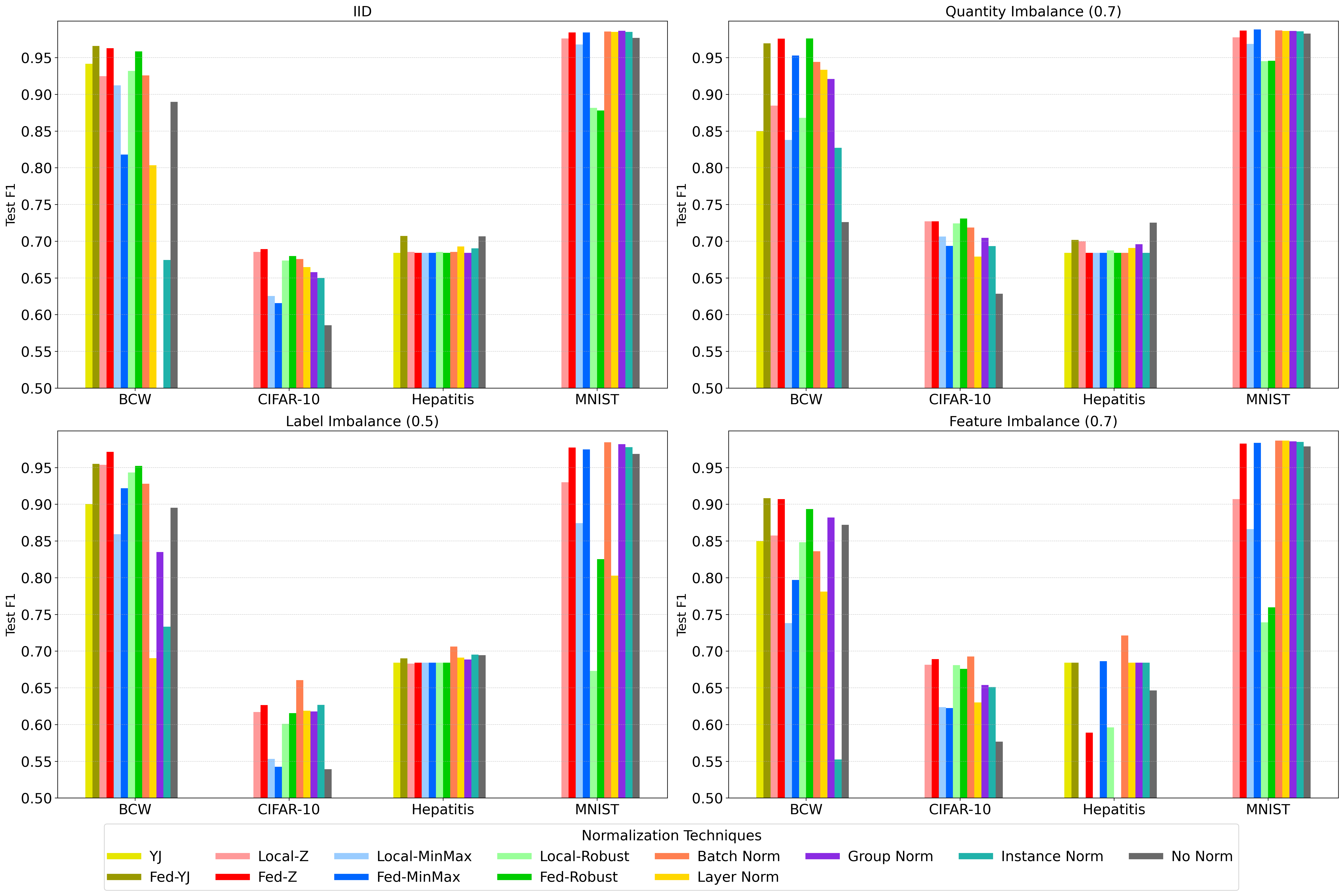}
\caption{Test F1 scores of various imbalance settings for 30 clients (50 for image datasets). Highly imbalanced (more heterogeneous) parameters are selected for each imbalance type. Only classification datasets (BCW, CIFAR-10, Hepatitis, MNIST) are included due to representation with F1 score metric. The x-axis shows the dataset names.}
\label{fig:grouped_plot}
\vspace{-1em}
\end{figure*}

\descr{Federated Normalization Outperforms Local Normalization.}
 
Figure~\ref{fig:avg_plot} displays the average performance of each normalization technique across all datasets. Federated normalization methods consistently outperform their local counterparts, supporting our key claim that aggregating normalization statistics globally leads to better generalization and robustness in FL—particularly under heterogeneous data distributions. Table~\ref{table:result_counts} summarizes the number of runs in which federated normalization achieved an average performance improvement between \%1 and \%8. Overall, federated normalization outperformed local normalization in 232 runs, while local normalization performed better in 59 runs. We use test F1 score for all datasets except Parkinson’s, where test $R^2$ is used; only runs with $R^2>0$ are included for meaningful comparisons.  Results were counted as ties if the improvement was below 0.5\% for all datasets.
\begin{table}[!t]
    \centering
    \begin{threeparttable}
    \renewcommand{\arraystretch}{1.2}
    \setlength{\tabcolsep}{6pt}

\caption{Comparison of Federated vs. Local Normalization Across Datasets. The “Federated” and “Local” columns report the number of runs each method outperformed the other, while “Mean Fed. Impr.” reports the average gain when federated normalization is superior.}
    
    \vspace{1mm}
    \begin{tabular*}{\linewidth}
    {@{\extracolsep{\fill}}lcccc}
        \toprule
        \textbf{Dataset} & \textbf{Federated} & \textbf{Local} & \textbf{Ties} & \textbf{Mean Fed. Impr.} \\
        \midrule
        BCW          & 71 & 4 & 9  & 0.0414 \\
        CIFAR10      & 16 & 15 & 32 & 0.0113 \\
        Hepatisis    & 33 & 21 & 30 & 0.0567 \\
        MNIST        & 45 & 2  & 16 & 0.0817 \\
        Parkinson's  & 67 & 17 & 0  & 0.0645* \\
        \midrule
        Total  & 232 & 59 & 87  &  \\
        \bottomrule
    \label{table:result_counts}
    \end{tabular*}
    \begin{tablenotes}
        \footnotesize
        \item[$*$] Mean Fed. Impr. for Parkinson's dataset calculated on test $R^2$ score, while others calculated on test F1 score. 
    \end{tablenotes}
    \end{threeparttable}
    \vspace{-1em}
\end{table}

Except for CIFAR10, results show notable gains in non-IID settings. CIFAR10’s smaller gap likely stems from its higher complexity, which lessens the effect of applied feature imbalance. Overall, local normalization, while effective centrally, struggles in FL due to client-specific biases. A key observation is that, on average, models perform better even \textit{without any normalization} than \textit{with local normalization} (see Figure~\ref{fig:avg_plot}). This aligns with the findings of Zhang et al.~\cite{healthcare_survey}, who argue that local normalization can amplify client-specific biases in FL. These patterns are even more evident when examining specific imbalance types, as discussed in the next subsection.

From Figure~\ref{fig:avg_plot} and Figure~\ref{fig:grouped_plot}, we also observe that activation layer normalizations—i.e., batch, layer, instance, and group normalization—perform comparably to various federated normalization techniques. However, it is important to note that activation layer normal
ization techniques are not universally applicable to all PPFL methods, particularly those that rely on FHE~\cite{wang2023batch,poseidon} and require certain reformulations~\cite{fhebnorm}. In privacy-critical federated applications, such techniques must be applied cautiously or augmented with privacy-preserving mechanisms—an effort that is often complex and not always straightforward. Thus, we focus on the federated normalization techniques in later observations.

\descr{Performance Under Imbalanced Distributions.}
Figure~\ref{fig:grouped_plot} shows the performance of all normalization techniques across IID, quantity, label, and feature imbalance settings with 30 clients (see Supplementary Material~\ref{sec:supFigures}, Figure~\ref{fig:grouped_plot_low_skew}, for additional variations). Under IID distributions, most techniques perform similarly, as expected. However, with imbalanced data, the gap between local and federated normalization becomes substantial.

\begin{figure}[!t]
\centering
\includegraphics[width=0.8\columnwidth]{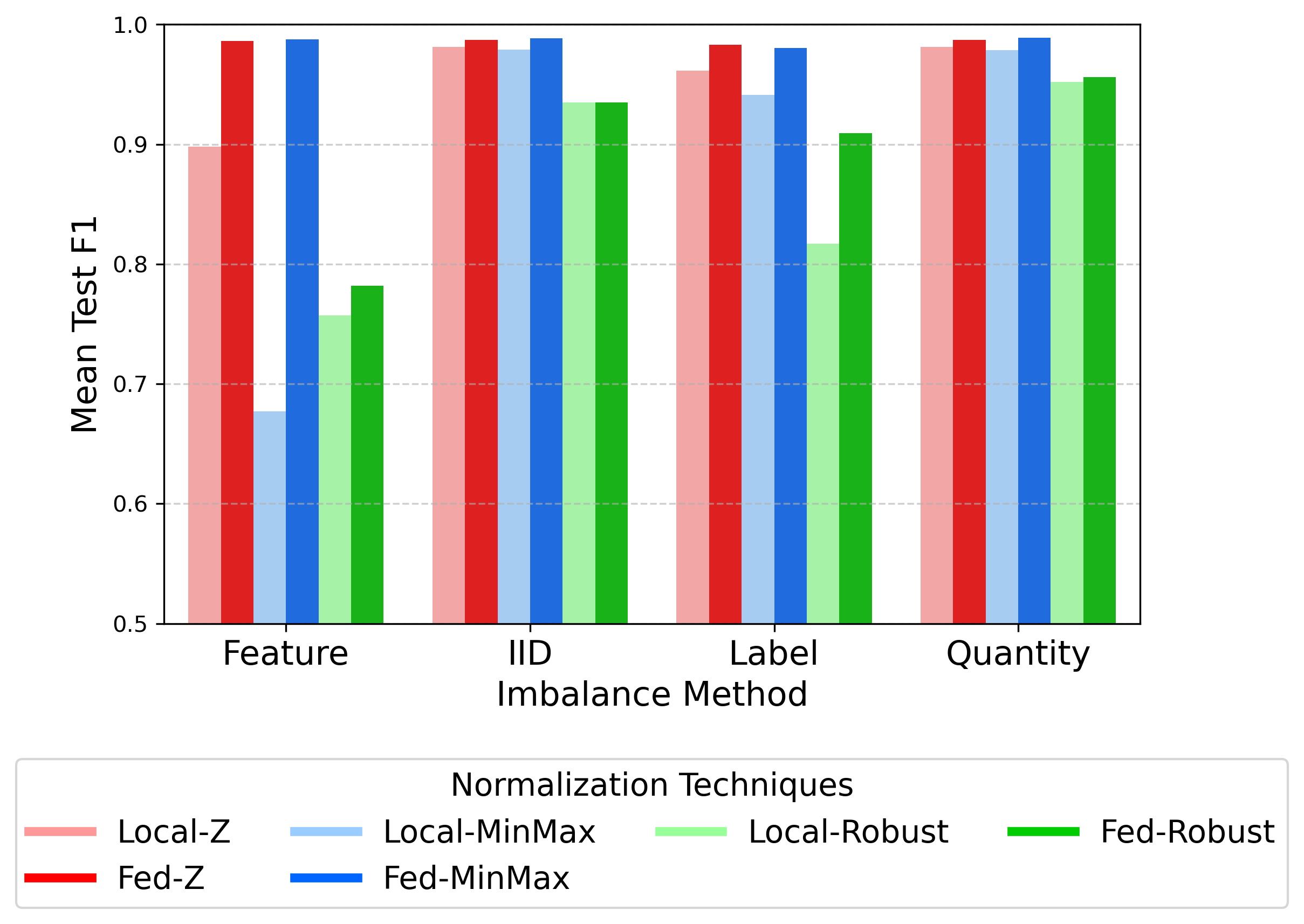}
\caption{Test F1 results of MNIST dataset averaged over client numbers to observe the impact of different heterogeneous scenarios. We observe that under the feature-imbalanced distribution, federated normalization outperforms local.}
\label{fig:mnist_skew}
\vspace{-1em}
\end{figure}

Federated techniques retain high F1 scores across all imbalance scenarios. In contrast, local normalization methods degrade significantly under non-IID settings, especially when clients exhibit non-overlapping feature ranges or label distributions. This trend is further reinforced by results on the MNIST dataset shown in Figure~\ref{fig:mnist_skew} and also for Hepatitis, Parkinson's and BCW datasets in Supplementary Material~\ref{sec:supFigures}, Figure~\ref{fig:hepatitis_skew}, ~\ref{fig:parkinson_skew} and ~\ref{fig:bcw_skew}. Although all normalization methods perform well under IID settings, federated normalization techniques continue to outperform local methods across all types of imbalance. Particularly under \textit{label} and \textit{feature} imbalance, federated methods maintain strong performance while local methods degrade sharply. In quantity imbalance, performance resembles IID settings, as clients with large data shares can dominate training, mimicking centralized learning. 

\begin{figure}[t]
\centering
\includegraphics[width=0.75\columnwidth]{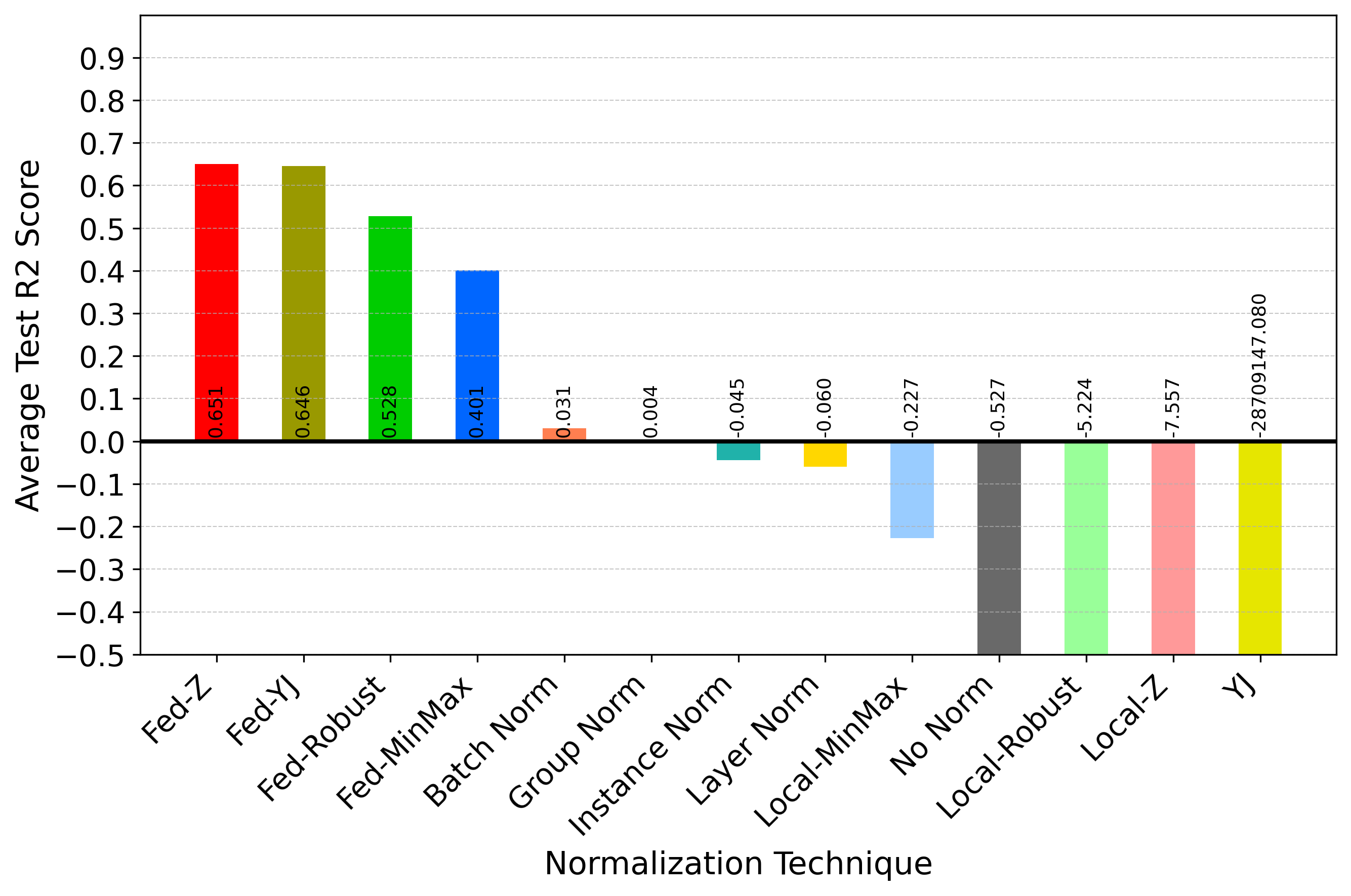}
\vspace{-1em}
\caption{Test $R^2$ scores of Parkinson's Monitoring dataset experiments under feature imbalance, averaged over normalization techniques used. Results are sorted according to $R^2$ scores.}
\label{fig:parkinson}
\vspace{-1em}
\end{figure}

In regression tasks, the gap between federated and local normalization is more pronounced. Under feature imbalance, local normalization often fails to converge, whereas federated normalization consistently yields strong performance (see Figure~\ref{fig:parkinson}).

\begin{figure}[t]
\centering
\includegraphics[width=0.75\columnwidth]
{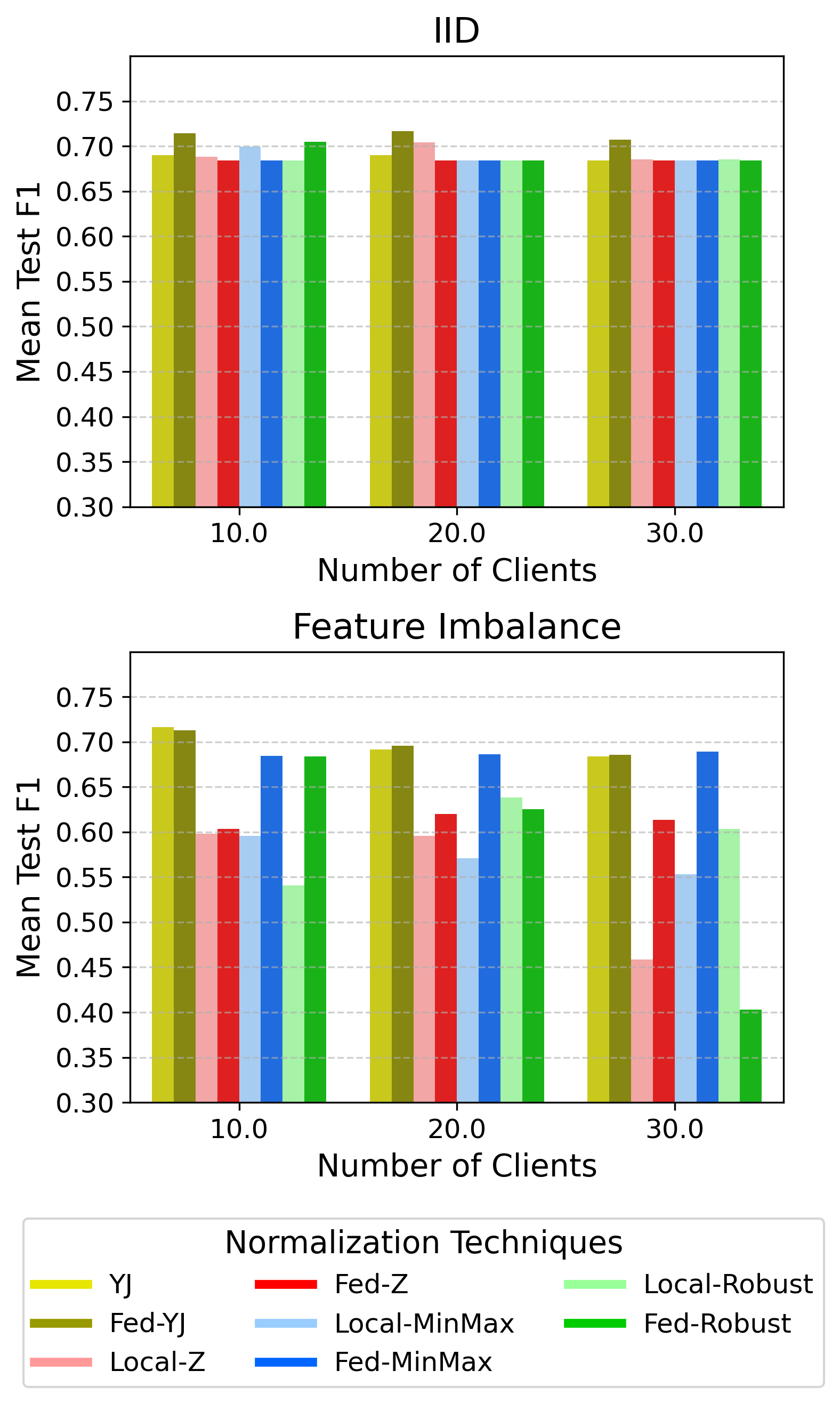}
\caption{Test F1 scores for Hepatitis dataset for feature imbalance \& IID comparison. In feature-imbalanced scenarios, increasing the number of clients amplifies the performance gap between local and federated normalization. Under IID settings, the results remain closely aligned regardless of the normalization method.}
\label{fig:hepatit}
\vspace{-1em}
\end{figure}

\descr{Impact of Client Count Under Feature Imbalance.}
To examine the impact of the number of clients, we provide Figure~\ref{fig:hepatit}, focusing on the Hepatitis dataset, comparing performance across IID and feature-imbalanced settings (see Supplementary Material~\ref{sec:supFigures}, Figure~\ref{fig:mnist_client} and ~\ref{fig:bcw_client} for MNIST and BCW datasets). While performance gap between federated and local normalizations remain stable under IID across varying client counts, feature imbalance reveals a mild decline in local normalization performance as client numbers grow. This suggests that smaller client datasets benefit more from global normalization parameters.

\subsection{Key Takeaways}
\begin{itemize}
    \item \textbf{Federated normalization techniques outperform local ones} on average, especially under feature and label imbalance, aligning with our thesis. Local normalization, in contrast, fails to align the feature distributions across clients, leading to less performance and instability.
    \item On average, \textbf{models without normalization outperform those using local normalization}, suggesting that local normalization may amplify client-specific biases.
    \item \textbf{Feature imbalance} introduces significant performance challenges; only robust federated methods remain stable.
    \item \textbf{Quantity imbalance} does not significantly affect performance and yields results similar to IID scenarios.
    \item \textbf{Activation layer normalizations} perform comparably to federated normalization but are challenging to integrate into PPFL frameworks due to privacy constraints.
\end{itemize}

\section{Privacy Preserving Federated Normalization}\label{sec:pp-normalization}
Our study in Section~\ref{sec:results} demonstrates that federated normalization outperforms other normalization techniques. However, sharing local parameters with the aggregation server, such as the minimum or maximum values of a client’s dataset, poses a privacy risk, particularly in heterogeneous FL environments. To mitigate this leakage, we propose novel algorithms that enable federated normalization under MHE. Our primary goal is to enable the execution of previously mentioned federated normalization techniques \textit{under encryption, ensuring that sensitive values remain encrypted throughout the process.}

We present privacy-preserving federated (PPF) protocols for z-score normalization, MinMax scaling, and robust scaling with MHE-CKKS. The computations for these protocols are executed on the FL aggregation server. Below, we describe the protocols for PPF normalization techniques. 
\par\noindent\textbf{Notations.} In the algorithms presented in this section, $\bigoplus_{p=1}^P$ denotes the homomorphic summation of elements, which is the homomorphic counterpart of $\sum_{p=1}^{P}$. Similarly, $\odot$ represents the homomorphic multiplication. We use $\langle X \rangle$ to denote the encrypted version of a plaintext $X$. Other homomorphic operations are self-explanatory based on their names. We define the core operations (e.g., encryption) in the Background section~\ref{mhebackground}. The \textsf{Inv($\langle m \rangle$)} operation represents the homomorphic inverse on encrypted message $\langle m \rangle$, which we rely on implementation in the Lattigo package~\cite{lattigo}. This implementation in Lattigo relies on the Goldschmidt division algorithm~\cite{homomorphicInverse, goldschmidt}. \textsf{Min($\langle m_1 \rangle$, $\langle m_2 \rangle$)}, \textsf{Max($\langle m_1 \rangle$, $\langle m_2 \rangle$)} stands for homomorphic minimum and maximum comparison operations, which we rely on the minimax approximation on the Lattigo package~\cite{lattigo}. The prefix C in operation names stands for Collective; for example, \textsf{CBootstrap($\langle m_1 \rangle$, \{$sk_p$\})} refers to the Collective Bootstrapping operation. The PPF Robust Scaling protocol leverages PPF MinMax Scaling and PPF $k$-th Element Calculation as \textsf{PPFMinMax()} and \textsf{PPF-Kth()}, respectively.

\subsection{PPF Z-Score Normalization}\label{sec:pp-zscore}

Federated z-score normalization requires computing the global mean ($\mu_g$) and variance ($\sigma_g^2$) per feature as if the data were pooled, without centralizing it. Our protocol reveals only the final outputs: the collaboratively computed mean ($\mu_g$) and variance ($\sigma_g^2$). We summarize the protocol in Algorithm~\ref{alg:federated_zscore} and provide detailed steps below.

Parties begin by generating secret keys and jointly establishing a public encryption key (Line 1, Algorithm~\ref{alg:federated_zscore}). Each party then computes the local summation \(S_p\), and the number of samples \(n_p\) for each feature in their dataset (Lines 2--7), forming arrays of size \(N\), where \(N\) is the total number of features. These arrays are encrypted using the collective public key—denoted as \(\langle S_p \rangle\) (local sums) and \(\langle n_p \rangle\) (local sample counts)—, with each value occupying a slot in the ciphertext, utilizing \(N\) slots per ciphertext (Line 8). The encrypted values are then sent to the server (Line 9), which aggregates them to compute \(\langle S_g \rangle\) (global sums) and \(\langle n_g \rangle\) (global sample counts), representing the feature-wise encrypted summation and sample count as if the data were centrally pooled (Lines 11--13).

Then, the server computes the inverse of $\langle n_g \rangle$ in the encrypted domain, and the parties apply collective bootstrapping to restore the level of the ciphertext (Lines 14--16), yielding $\langle n_g^{-1} \rangle$. The calculated inverse and $\langle S_g \rangle$ are then homomorphically multiplied to compute $\langle \mu_g \rangle$ (Line 17). Finally, $\langle \mu_g \rangle$ is collectively decrypted, making the global mean ($\mu_g$) available to all parties (Line 18). Upon decryption, our protocol proceeds to calculate the global variances. Each party subtracts $\mu_g$ from their local feature values and squares the results (Lines 20--23). The sums of these squared differences for each feature are computed locally, resulting in arrays $S_{p,\text{sq}}$. These arrays are encrypted, denoted as $\langle S_{p,\text{sq}} \rangle$, and sent to the server (Lines 24--25). The server aggregates the encrypted values to compute $\langle S_{g,\text{sq}} \rangle$ (global squared difference sums) (Line 28). Subsequently, the server homomorphically multiplies $\langle S_{g,\text{sq}} \rangle$ with $\langle n_g^{-1} \rangle$ to compute $\langle \sigma_g^{2} \rangle$ (Line 29). $\langle \sigma_g^{2} \rangle$ is then collectively decrypted to obtain $\sigma_g^2$ (Line 31). Finally, each client receives \(\mu_g\) and \(\sigma_g^2\), which are then used locally to apply z-score normalization to their data (Line 32).

\begin{algorithm}[t]
\caption{PPF Z-Score Normalization Protocol}
\label{alg:federated_zscore}
\small
\begin{algorithmic}[1]

\STATE \textbf{Initialization:} $\{sk_p\}, pk \gets \textsf{KeyGen($\lambda$)}$

\STATE \textbf{Compute Local Summations (Client Side):}
\FOR{each party $p$ \textbf{in} $\{1, \ldots, P\}$}
    \FOR{each feature $j$ \textbf{in} $\{1, \ldots, N\}$}
        \STATE $n_p[j] \gets$ number of samples for feature $j$
        \STATE $S_p[j] \gets \sum_{i=1}^{n_p[j]} D_p[j][i]$
    \ENDFOR
    \STATE $\langle S_p \rangle$, $\langle n_p \rangle$ $\gets$ \textsf{Encrypt($S_p$, $pk$)}, \textsf{Encrypt($n_p$, $pk$)}
    \STATE Send $\langle S_p \rangle$ and $\langle n_p \rangle$ to the server
\ENDFOR

\STATE \textbf{Global Summations (Server side):}
\STATE $\langle S_g \rangle \gets \bigoplus_{p=1}^P \langle S_p \rangle$ 
\STATE $\langle n_g \rangle \gets \bigoplus_{p=1}^P \langle n_p \rangle$ 

\STATE \textbf{Compute Global Means (Server side):}
\STATE $\langle n_g^{-1} \rangle \gets$ \textsf{Inv($\langle n_g \rangle$)}
\STATE $\langle n_g^{-1} \rangle \gets \textsf{CBootstrap}(\langle n_g^{-1} \rangle, \{sk_p\})$
\STATE $\langle \mu_g \rangle \gets \langle S_g \rangle \odot \langle n_g^{-1} \rangle$ 
\STATE $\mu_g$ $\gets$ \textsf{CDecrypt($\langle \mu_g \rangle$, \{$sk_p$\})}

\STATE \textbf{Compute Local Summation of Squared Differences (Client Side):}
\FOR{each party $p$ \textbf{in} $\{1, \ldots, P\}$}
    \FOR{each feature $j$ \textbf{in} $\{1, \ldots, N\}$}
        \STATE $S_{p,\text{sq}}[j] \gets \sum_{i=1}^{n_p[j]} (D_p[j][i] - \mu_g)^2$
    \ENDFOR
    \STATE $\langle S_{p,\text{sq}} \rangle$ $\gets$ \textsf{Encrypt($S_{p,\text{sq}}$, $pk$)}
    \STATE Send $S_{p,\text{sq}}^*$ to the server
\ENDFOR

\STATE \textbf{Compute Global Variances (Server side):}
\STATE $\langle S_{g,\text{sq}} \rangle \gets \bigoplus_{p=1}^P \langle S_{p,\text{sq}} \rangle$
\STATE $\langle \sigma_g^{2} \rangle \gets \langle S_{g,\text{sq}} \rangle \odot \langle n_g^{-1} \rangle$
\STATE \textbf{Decryption and Local Normalization (Client Side):}
\STATE $\sigma_g^2$ $\gets$ \textsf{CDecrypt($\langle \sigma_g^{2} \rangle$, \{$sk_p$\})}

\STATE Parties apply z-score normalization locally using $\mu_g$, $\sigma_g^2$.
\end{algorithmic}
\end{algorithm}
\subsection{PPF MinMax Scaling}

Federated MinMax scaling requires global feature-wise minima and maxima, as if the data were pooled. Our protocol reveals only the global minimum ($X_{\min,g}$) and maximum ($X_{\max,g}$) for each feature. To preserve privacy, the protocol uses homomorphic comparison functions, which operate on inputs within the range $[-1,1]$. These functions securely compute element-wise minima or maxima across ciphertexts. Since this algorithm closely resembles PPF Z-score normalization, with the main distinction being the evaluation of minimum and maximum values under encryption, we provide the algorithm and its detailed description in Supplementary Material~\ref{sec:supminmax}.

\subsection{PPF Robust Scaling}\label{subsec:federated_robust}

Applying PPF Robust Scaling requires computing the median, 25th percentile, and 75th percentile values of each feature as if the datasets of all parties were pooled. The core algorithm is a general-purpose method for computing the $k$-th ranked elements of features in FL while preserving privacy.

The PPF Robust Scaling protocol also begins with key generation (Line 1 of Algorithm~\ref{alg:federated_robust}). Next, each party computes the number of samples ($n_p$) for every feature in its local dataset (Lines 3--6) and encrypts these counts ($\langle n_p \rangle$) before sending them to the server (Lines 7--8). On the server side, these encrypted counts are aggregated (Lines 10--11) to obtain global sample counts ($\langle n_g \rangle$). The parties then collectively decrypt $\langle n_g \rangle$ to obtain the global sample counts $n_g$ (Lines 12).

Using the global sample counts, the indices corresponding to the median, 25th percentile, and 75th percentile —stored as arrays $K_{Q1}$, $K_{Q2}$, and $K_{Q3}$ along with boolean arrays $K_{b,Q1}$, $K_{b,Q2}$, and $K_{b,Q3}$ respectively— are calculated for each feature (Lines 13--17). Boolean arrays ($K_{b,Q*}$) indicate whether the percentile for feature $j$ is at an exact index or requires interpolation (e.g., median of a length-10 array). The protocol then computes global feature-wise minima and maxima using PPF MinMax Scaling (Lines 18-19). Then, for each percentile (denoted as $Q*$ in \{Q1, Q2, Q3\}), the server calls the PPF $k$-th Element Calculation protocol (Algorithm~\ref{alg:secure_kth}). This algorithm performs a binary search over a floating-point range to find the value meeting the rank condition. The search range is bounded by global min-max values, and precision is controlled by a threshold $\epsilon$ to ensure termination. Once the target percentiles are identified, each party applies robust scaling locally using these values (Lines 25–26).

Below, we introduce our PPF $k$-th Ranked Element Calculation algorithm which we build upon the algorithm proposed in~\cite{median}. Our novel approach computes the $k$-th ranked element using homomorphic encryption based on the CKKS scheme, enabling privacy preservation while supporting floating-point inputs for the first time in this context. This contribution represents a significant advancement not only for federated normalization but also broadly for PPF protocols.

\begin{algorithm}[t]
\caption{PPF Robust Scaling Protocol}
\label{alg:federated_robust}
\small
\begin{algorithmic}[1]

\STATE \textbf{Initialization:} $\{sk_p\}, pk \gets \textsf{KeyGen($\lambda$)}$

\STATE \textbf{Compute Local Sample Counts (Client Side):}
\FOR{each party $p$ \textbf{in} $\{1, \ldots, P\}$}
    \FOR{each feature $j$ \textbf{in} $\{1, \ldots, N\}$}
        \STATE $n_p[j] \gets$ number of samples for feature $j$
    \ENDFOR
    \STATE $\langle n_p \rangle \gets$ \textsf{Encrypt($n_p$, $pk$)}
    \STATE Send $\langle n_p \rangle$ to the server
\ENDFOR

\STATE \textbf{Global Sample Counts (Server Side):}
\STATE $\langle n_g \rangle \gets \bigoplus_{p=1}^P \langle n_p \rangle$
\STATE $n_g \gets$ \textsf{CDecrypt($\langle n_g \rangle$, \{$sk_p$\})}

\STATE \textbf{Determine Percentile Indices (Server Side):}

\FOR{each feature $j$ in $\{1, \ldots, N\}$}
    \STATE Compute the 25th, 50th, and 75th percentiles of $n_g[j]$ as $K_{Q1}[j]$, $K_{Q2}[j]$, $K_{Q3}[j]$
    \STATE For each $K_{Q*}[j]$, set $K_{b,Q*}[j] \gets \textbf{true}$ if given index is exact; else set to \textbf{false}
\ENDFOR

\STATE \textbf{Compute Global Min-Max:}

\STATE $X_{\min,g}$, $X_{\max,g}$ $\gets$ \textsf{PPFMinMax($D_1, D_2, \ldots, D_P$)} 

\STATE \textbf{Compute Percentiles (Server Side):}
\FOR{each percentile $Q*$ \textbf{in} \{Q1, Q2, Q3\}}
    \STATE Set $K \gets K_{Q*}$, $K_b \gets K_{b,Q*}$
    \STATE $Q*_g \gets \textsf{PPF-Kth(}X_{\min,g}, X_{\max,g}, K_q, K_{b,q}, n_g, \epsilon\textsf{)}$ 
\ENDFOR

\STATE \textbf{Local Normalization (Client Side):}
\STATE Parties locally apply robust scaling using $Q1_g$, $Q2_g$, $Q3_g$
\end{algorithmic}
\end{algorithm}

\descr{PPF $k$-th Ranked Element Calculation}
Our proposed protocol begins by initializing an empty boolean array $B$ of size $N$ to track whether the $k$-th element for each feature has been found (Lines 1--2 of Algorithm~\ref{alg:secure_kth}). Then, for each feature, the lower ($\min_j$) and upper ($\max_j$) bounds of the search range are set using the global minimum and maximum values (Line 3).

\begin{algorithm}[!t]
\caption{PPF $k$-th Element Calculation (\textsf{PPF-Kth()})}
\label{alg:secure_kth}
\small
\begin{algorithmic}[1]

\STATE \textbf{Initialization:}
\STATE Initialize $B[j] \gets \text{False}$ $\forall j$, $Q_g[j] \gets 0$ $\forall j$
\STATE $\text{min}_j \gets X_{\min,g}[j]$, $\text{max}_j \gets X_{\max,g}[j]$ $\forall j$

\REPEAT
\STATE \textbf{Compute Midpoints (Server Side):}
\FOR{each feature $j$ \textbf{in} $\{1, \ldots, N\}$}
    \STATE $M[j] \gets (\text{min}_j + \text{max}_j)/2$
\ENDFOR
\STATE Send $M$ to the parties.

\STATE \textbf{Count Smaller and Greater Values (Client Side):}
\FOR{each party $p$ \textbf{in} $\{1, \ldots, P\}$}
    \FOR{each feature $j$ \textbf{in} $\{1, \ldots, N\}$}
        \STATE $Lcount_p[j] \gets \sum_{i=1}^{|D_p[j]|} \mathbf{1}(D_p[j][i] \leq M[j])$
        \STATE $Gcount_p[j] \gets \sum_{i=1}^{|D_p[j]|} \mathbf{1}(D_p[j][i] > M[j])$
    \ENDFOR
    \STATE $\langle Lcount_p \rangle \gets \textsf{Encrypt(}Lcount_p, pk\textsf{)}$ 
    \STATE $\langle Gcount_p \rangle \gets \textsf{Encrypt(}Gcount_p, pk\textsf{)}$
    \STATE Send $\langle Lcount_p \rangle$, $\langle Gcount_p \rangle$ to server.
\ENDFOR

\STATE \textbf{Aggregate Counts (Server Side):}
\STATE $\langle Lcount_g \rangle \gets \bigoplus_{p=1}^P \langle Lcount_p \rangle$
\STATE $\langle Gcount_g \rangle \gets \bigoplus_{p=1}^P \langle Gcount_p \rangle$
\STATE $Lcount_g \gets \textsf{CDecrypt(}\langle Lcount_g \rangle, \{sk_p\}\textsf{)}$ 
\STATE $Gcount_g \gets \textsf{CDecrypt(}\langle Gcount_g \rangle, \{sk_p\}\textsf{)}$

\STATE \textbf{Update Bounds and Results (Server Side):}
\FOR{each feature $j$ \textbf{in} $\{1, \ldots, N\}$}
    \IF{not $B[j]$}
        \IF{$K_b[j]$}
            \IF{$Lcount_g[j] \leq K[j] - 1$ \textbf{and} $Gcount_g[j] \leq n_g[j] - K[j]$}
                \STATE $Q_g[j] \gets M[j]$, $B[j] \gets \text{True}$
                \STATE continue
                \ENDIF
        \ELSE
            \IF{$Lcount_g[j] \leq K[j]$ \textbf{and} $Gcount_g[j] \leq n_g[j] - K[j]$}
                \STATE $Q_g[j] \gets M[j]$, $B[j] \gets \text{True}$
                \STATE continue
            \ENDIF
        \ENDIF
        \STATE \textbf{Adjust the search range:}
        \IF{$Lcount_g[j] \geq K[j]$}
            \STATE $\text{max}_j \gets M[j]$
        \ELSE
            \STATE $\text{min}_j \gets M[j]$
        \ENDIF
        \STATE \textbf{Check if reached precision limit:}
        \IF{$\text{max}_j - \text{min}_j \leq \epsilon$}
            \STATE $Q_g[j] \gets (\text{min}_j + \text{max}_j)/2$, $B[j] \gets \text{True}$
            \STATE continue
        \ENDIF

    \ENDIF
\ENDFOR
\UNTIL{$\forall j$, $B[j] = \text{True}$}
\end{algorithmic}
\end{algorithm}


In each iteration (starting at Line 4), the server computes the midpoint $M[j]$ of the current search interval for each feature $j$ (Lines 5--8). These midpoints are sent to the parties (Line 9), which locally count the number of data values smaller ($Lcount_p$) than or greater ($Gcount_p$) than the midpoint (Lines 10--15). Each party encrypts the resulting counts as $\langle Lcount_p \rangle$ and $\langle Gcount_p \rangle$ and sends to the server (Lines 16--18). 

On the server side, the encrypted counts from all parties are aggregated and then collectively decrypted to obtain the global counts (Lines 20--24). Next, the server checks whether the current midpoint satisfies the rank condition by comparing the counts to the target index $K[j]$ (Lines 28--38). If the condition is met, the $k$-th element is recorded (Lines 30, 35) and the corresponding boolean flag is set to \textit{true}. 

If the condition is not met, the algorithm adjusts the search range (Lines 39--44): If $Lcount_{g}[j] \geq K[j]$, $\max_j$ is updated to $M[j]$; otherwise, $\min_j$ is updated to $M[j]$. Additionally, if the updated search range is smaller than the precision threshold $\epsilon$, the algorithm finalizes the value as the average of the current $\min_j$ and $\max_j$ (Lines 45--49). This iterative process continues until all features have their $k$-th element computed (Line 52).

\section{Evaluation of PPF Normalization}

We evaluate the proposed PPF normalization protocols through theoretical and empirical analyses, detailing complexity via dominant operations and benchmarking performance. 
\subsection{Theoretical Analysis}\label{sec:theoreticalAnalysis}

In this section, we present the theoretical worst-case time and communication complexity analysis of the proposed protocols. For clarity and ease of interpretation, we express time complexity in terms of the time required for each dominant operation. Specifically, we denote the execution time of each fundamental operation by $T_{operation}$ —for example, $T_{encrypt}$ represents the encryption time. Later, we provide microbenchmark measurements of these operations to facilitate interpretation and extrapolation in the Empirical Analysis section~\ref{sec:empiricalAnalysis} under settings with varying number of parties.

\begin{table*}[!t]
  \centering
 
 \caption{Theoretical Time and Communication Complexities of PPF Protocols.}
  \setlength{\tabcolsep}{4pt}
  \renewcommand{\arraystretch}{1.2}
  \begin{tabular}{|c|c|c|}
    \hline
    \textbf{Protocol} & \textbf{Time Complexity}  & \textbf{Communication Complexity} \\ \hline
    PPF Z-Score Normalization & $3PT_{sum} + T_{cbootstrap} + T_{inv} + 2T_{mul}$ & $ 3P|c| + 3S_{cdecrypt} + \mathcal{K}_{inv} S_{cbootstrap}$ \\ \hline
    PPF MinMax Scaling ($\gamma$) & $4PT_{mul} + PT_{min} + PT_{max} + 2PT_{cbootstrap}$ & $2P|c| + 2S_{cdecrypt} + 2P\mathcal{K}_{mm}S_{cbootstrap}$ \\ \hline
    PPF Robust Scaling & $PT_{sum} + T_\gamma + 3T_\beta $ & $P|c| + S_{cdecrypt} + S_\gamma + 3S_\beta$ \\ \hline
    PPF $k$-th Ranked Element Calculation ($\beta$) & $\mathcal{X} \times (2PT_{sum} + 2T_{cdecrypt} + 2T_{encrpyt})$ & $\mathcal{X}  \times (2P|c| + 2S_{cdecrypt} + P|p|)$ \\ \hline
  \end{tabular}

  \label{tab:theoretical_analysis}
\end{table*}

Note that the PPF Robust Scaling protocol incorporates the PPF MinMax Scaling and the PPF $k$-th Ranked Element Calculation protocols. We denote the time complexities of these protocols as \(T_\gamma\) and \(T_\beta\), respectively, and use these to represent the overall time complexity of the PPF Robust Scaling protocol. The parameter \(\mathcal{X}\) appearing in both the time and communication complexity analysis for the PPF $k$-th Ranked Element Calculation protocol in Table~\ref{tab:theoretical_analysis} is defines as:
\[
\mathcal{X} = \log\left(\frac{\arg\max_{j}(max_j - min_j)}{\epsilon}\right),
\]
where \(\arg\max_{j}(max_j - min_j)\) denotes the search range for the feature \(j\) that has the maximum search range for the $k$-th element calculation protocol. The maximum search range among features determines the worst-case time and communication complexity, as the protocol must wait until computations for all features are completed in parallel.

For the theoretical time complexity representations, we consider only the dominant terms, i.e., the time-consuming repetitive homomorphic operations. One-time operations, such as key generation, encryption, or client-side plaintext operations, are omitted as their impact is negligible.

We present the communication complexities in Table~\ref{tab:theoretical_analysis} in terms of total bits transferred. In these representations, \(P\) denotes the number of parties, while \(|c|\) and \(|p|\) denote the size of a ciphertext and a plaintext, respectively. Additional variables, denoted by the letter \(S\), are used to simplify the presentation. Specifically, \(S_\gamma\) and \(S_\beta\) denote the communication complexities of the PPF MinMax Scaling and the PPF $k$-th Ranked Element Calculation protocols.

Furthermore, \(S_{cdecrypt}\) and \(S_{cbootstrap}\) represent the communication complexities of the collective decryption and collective bootstrapping operations, respectively, both with a predefined worst-case complexity of
$S_{cdecrypt} =  S_{cbootstrap} = (P-1)|c|$. The collective bootstrapping operation is the dominant factor in the overall communication complexities, as it is executed several times and requires the participation of all parties. Since the homomorphic min, max, and inverse operations incorporate the collective bootstrapping operation, we define the terms \(\mathcal{K}_{inv}\) and \(\mathcal{K}_{mm}\) as multipliers for \(S_{cbootstrap}\) in Table~\ref{tab:theoretical_analysis}.

The PPF Z-Score and PPF MinMax protocols rely on homomorphic inverse, minimum, and maximum operations, and the communication costs associated with these operations depend on \(\mathcal{K}_{inv}\) and \(\mathcal{K}_{mm}\). The values of \(\mathcal{K}_{inv}\) and \(\mathcal{K}_{mm}\) are determined by the chosen security parameters and, for \(\mathcal{K}_{inv}\), also by the acceptable input value range of homomorphic inverse operation. For example, in a 10-party setting with a security parameter of \(\log N = 15\), we have \(\mathcal{K}_{mm} = 1\) for the homomorphic minimum and maximum operations, and \(\mathcal{K}_{inv} = 13\) for the homomorphic inverse operation where maximum absolute input value is $|2^{20}|$ (i.e., it can take inputs in the range $-2^{20}$ to $2^{20}$). Our analysis shows that PPF MinMax scaling is the most time- and communication-intensive, due to bootstrapping operations that scale with the number of parties. This, in turn, increases the overall cost of PPF Robust Scaling.

\subsection{Empirical Analysis}\label{sec:empiricalAnalysis}

\begin{table}[!t]
    \centering
    \caption{Microbenchmarks with 10 clients}
    \label{tab:micro_10clients2}
    \renewcommand{\arraystretch}{1.2}
    \begin{tabular}{@{}lc@{}}
        \toprule
        \textbf{Operation Name} & \textbf{Time (ms)} \\
        \midrule
        \multicolumn{2}{c}{\textbf{Common Operations}} \\
        \midrule
        Secret Key Generation ($T_{skgen}$) & 9.67 \\
        Collective Public Key Generation ($T_{pkgen}$) & 138.71 \\
        Collective Relinearization Key Generation ($T_{relkgen}$) & 5507.03 \\
        Encryption ($T_{encrypt}$) & 59.28 \\
        Collective Decryption ($T_{decrypt}$) & 536.63 \\
        \cmidrule(lr){1-2}
        \multicolumn{2}{c}{\textbf{PPF Z-Score Normalization}} \\
        \cmidrule(lr){1-2}
        Homomorphic Multiplication ($T_{mul}$) & 173.77 \\
        Homomorphic Summation ($T_{sum}$) & 1.73 \\
        Homomorphic Inverse ($T_{inv}$) & 30135.78 \\
        Collective Bootstrapping ($T_{bootstrap}$) & 825.34 \\
        \cmidrule(lr){1-2}
        \multicolumn{2}{c}{\textbf{PPF MinMax Scaling}} \\
        \cmidrule(lr){1-2}
        Collective Galois Keys generation ($T_{gkgen}$) & 1908.93 \\
        Homomorphic Multiplication ($T_{mul}$) & 173.77 \\
        Homomorphic Inverse ($T_{inv}$) & 30135.78 \\
        Homomorphic Minimum Comparison ($T_{min}$) & 12255.31 \\
        Homomorphic Maximum Comparison ($T_{max}$) & 12790.85 \\
        Collective Bootstrapping ($T_{bootstrap}$) & 825.34 \\
        \cmidrule(lr){1-2}
        \multicolumn{2}{c}{\textbf{PPF Robust Scaling}} \\
        \cmidrule(lr){1-2}
        Homomorphic Summation ($T_{sum}$) & 1.73 \\
        \bottomrule
    \end{tabular}
    \vspace{-1em}
\end{table}

\begin{table}[!t]
\caption{Total Runtimes of PPF Normalizations for 10 Parties. Precision value ($\epsilon$) for PPF Robust Scaling is taken as $1e^{-4}$.}
\label{tab:totalruntime}
\centering
\begin{tabular}{lc}
\toprule
\textbf{Protocol} & \textbf{Time (sec)} \\
\midrule
PPF Z-Score Normalization & 44 \\
PPF MinMax Scaling & 253 \\
PPF Robust Scaling & 368 \\
\bottomrule
\end{tabular}
\vspace{-1em}
\end{table}

\subsubsection{Implementation Details}

We evaluate the empirical runtime on a system with a 12\textsuperscript{th} Gen Intel\textsuperscript{\textregistered} Core\textsuperscript{\texttrademark} i5-12400F processor (2.50~GHz base frequency, 6 physical cores, 12 logical threads) and 32~GB of DDR4 RAM operating at 3600~MT/s. Unless otherwise specified, the security parameters for the CKKS scheme are set to $\lambda=128$-bit security with a ring size of $\log N = 15$, and the maximum absolute input value for the homomorphic inverse operation is set to $|2^{30}|$.

\subsubsection{Microbenchmarks}

Table~\ref{tab:micro_10clients2} presents benchmark results for micro-level homomorphic operations with 10 clients. We also present the same benchmarks for 20, 30, and 50 party settings in Supplementary Material~\ref{sec:supFigures}, Tables~\ref{tab:micro_20clients2}, ~\ref{tab:micro_30clients2}, and ~\ref{tab:micro_50clients2}. Tables categorize operations by protocol, with common steps such as encryption and collective decryption listed at the top.

Some operations—namely, $T_{skgen}$, $T_{encrypt}$, $T_{sum}$, and $T_{mul}$—are independent of the number of parties, as they do not require any collaboration. Therefore, the empirical runtimes of these operations remain constant across different party settings. In contrast, other operations involve inter-party collaboration and scale linearly with the number of parties (see Supplementary Material~\ref{sec:supFigures}). Reported runtimes exclude communication overhead, which varies with bandwidth, party location, and network latency.

\begin{table}[!t]
\caption{Precision loss of PPF normalization techniques. \textbf{Large Valued} indicates datasets with high-magnitude floats, while \textbf{Small Valued} refers to values near zero.}
\label{tab:precision_results}
\centering
\footnotesize
\sisetup{
  table-format=1.2e-2,
  exponent-product=\ensuremath{\times},
  tight-spacing=true,
  table-number-alignment=center
}
\begin{tabular}{@{} ll S[table-format=1.2e-2] S[table-format=1.2e-2] @{}}
\toprule
\textbf{Norm. Type} & \textbf{Parameters} & \textbf{Large Valued} & \textbf{Small Valued} \\
\midrule
\multirow{2}{*}{PPF Z-Score} 
  & Mean     & 1.80e-5 & 2.32e-6 \\
  & Variance & 1.80e-5 & 2.35e-5 \\
\cmidrule(r){1-4}
\multirow{2}{*}{PPF MinMax} 
  & Min      & 2.92e-9 & 1.78e-4 \\
  & Max      & 1.27e-8 & 4.64e-4 \\
\cmidrule(r){1-4}
\multirow{2}{*}{PPF Robust} 
  & Median ($\epsilon\!=\!10^{-6}$) & 1.45e-9 & 3.69e-7 \\
  & Median ($\epsilon\!=\!10^{-3}$) & 9.13e-7 & 3.07e-4 \\
\bottomrule
\end{tabular}
\vspace{-1em}
\end{table}

From all microbenchmark tables, we observe that certain operations dominate the computational cost for specific protocols—for instance, $T_{inv}$, $T_{min}$, and $T_{max}$. These operations are inherently dependent on $T_{bootstrap}$, as discussed in Section~\ref{sec:theoreticalAnalysis}. Additionally, the empirical benchmark values provided in microbenchmark tables, in combination with the theoretical analysis in Table~\ref{tab:theoretical_analysis}, can be used to estimate the runtime for new use cases under different settings. The following section demonstrates an example experimental total runtime for a 10-party setup.

\subsubsection{Total Runtime}

The computational runtimes of the proposed protocols for a 10-party setting are presented in Table~\ref{tab:totalruntime}. It is important to note that these runtimes do not account for communication overhead or client-side plaintext operations—such as calculating the mean of a local dataset—as these are negligible compared to computationally intensive ciphertext operations. Since the protocols are executed only once during the data preprocessing stage for PPML and require only a few minutes, as indicated in Table~\ref{tab:totalruntime}, they are practical and readily applicable to real-world FL scenarios. 

We further investigate the impact of the precision value ($\epsilon$) on the PPF $k$-th element algorithm (see Supplementary Material~\ref{sec:supFigures}, Figure~\ref{fig:epsilon_impact}). We observe a linear increase in runtime as $\epsilon$ decreases exponentially, which supports the theoretical analysis in Table~\ref{tab:theoretical_analysis}. Additionally, we observe a linear relationship between the number of clients and the runtime, indicating that the protocol scales well and remains practical for large deployments, including cross-device FL.

\subsubsection{Precision}
While CKKS encryption allows us to securely compute normalization parameters, it introduces a slight precision loss due to approximate encoding and noise accumulation inherent in fixed-point arithmetic on encrypted data. Table~\ref{tab:precision_results} summarizes the precision loss for each privacy-preserving normalization technique compared to computations on plaintext data. We evaluate using a 10-party scenario and consider two cases: one involving small floating-point values (close to zero), and another involving large floating-point values. We observe that as the values move away from zero (i.e., become larger), precision loss decreases. For example, for PPF MinMax Scaling, in a small-valued (all values close to 0) dataset, we can calculate the minimum value of the dataset with only a $1.73 \times 10^{-4}$ relative error, while for a larger-valued dataset, the relative error even decreases to $2.915 \times 10^{-9}$. For the median calculation in PPF Robust Scaling, precision depends on the $\epsilon$ value defined by the parties. We observe a direct correlation between the defined $\epsilon$ value and the precision loss, as shown in Table~\ref{tab:precision_results}.

\section{Conclusion}
In this work, we addressed the critical challenge of data normalization in federated learning, particularly under non-IID data settings. We proposed and evaluated federated normalization techniques that simulate pooled normalization, demonstrating their superiority over traditional local normalization methods. To further enhance privacy, we introduced novel protocols leveraging MHE for z-score normalization, MinMax scaling, and robust scaling. Notably, we also introduced the PPF $k$-th Element Calculation protocol, a general-purpose, privacy-preserving primitive that enables robust scaling and has potential applications beyond normalization. Collectively, our contributions advance the field of privacy-preserving federated learning by providing effective, secure, and practical normalization strategies suitable for real-world deployments.

\section*{Acknowledgements}
This work is supported by TUBITAK Career Grant \#124E091. The authors used ChatGPT-4o~\cite{openai2024gpt4o} to enhance the manuscript by shortening sentences, correcting typos, and grammar.

\bibliographystyle{IEEEtran}
\bibliography{example_paper}

\begin{thebibliography}{10}
\providecommand{\url}[1]{#1}
\csname url@samestyle\endcsname
\providecommand{\newblock}{\relax}
\providecommand{\bibinfo}[2]{#2}
\providecommand{\BIBentrySTDinterwordspacing}{\spaceskip=0pt\relax}
\providecommand{\BIBentryALTinterwordstretchfactor}{4}
\providecommand{\BIBentryALTinterwordspacing}{\spaceskip=\fontdimen2\font plus
\BIBentryALTinterwordstretchfactor\fontdimen3\font minus \fontdimen4\font\relax}
\providecommand{\BIBforeignlanguage}[2]{{%
\expandafter\ifx\csname l@#1\endcsname\relax
\typeout{** WARNING: IEEEtran.bst: No hyphenation pattern has been}%
\typeout{** loaded for the language `#1'. Using the pattern for}%
\typeout{** the default language instead.}%
\else
\language=\csname l@#1\endcsname
\fi
#2}}
\providecommand{\BIBdecl}{\relax}
\BIBdecl

\bibitem{Konency2016fed}
J.~Kone{\v{c}}n{\`y}, H.~B. McMahan, D.~Ramage, and P.~Richt{\'a}rik, ``Federated optimization: Distributed machine learning for on-device intelligence,'' \emph{CoRR}, vol. abs:1610.02527, 2016.

\bibitem{federatedLearning1}
B.~McMahan, E.~Moore, D.~Ramage, S.~Hampson, and B.~A. y~Arcas, ``Communication-efficient learning of deep networks from decentralized data,'' in \emph{Artificial intelligence and statistics}.\hskip 1em plus 0.5em minus 0.4em\relax PMLR, 2017, pp. 1273--1282.

\bibitem{Hitaj2017}
\BIBentryALTinterwordspacing
B.~Hitaj, G.~Ateniese, and F.~Perez-Cruz, ``Deep models under the {GAN}: Information leakage from collaborative deep learning,'' in \emph{Proceedings of the 2017 ACM SIGSAC Conference on Computer and Communications Security}, ser. CCS ’17.\hskip 1em plus 0.5em minus 0.4em\relax New York, NY, USA: Association for Computing Machinery, 2017, p. 603–618. [Online]. Available: \url{https://doi.org/10.1145/3133956.3134012}
\BIBentrySTDinterwordspacing

\bibitem{Wang2019}
Z.~{Wang}, M.~{Song}, Z.~{Zhang}, Y.~{Song}, Q.~{Wang}, and H.~{Qi}, ``Beyond inferring class representatives: User-level privacy leakage from federated learning,'' in \emph{IEEE INFOCOM 2019 - IEEE Conference on Computer Communications}, 2019, pp. 2512--2520.

\bibitem{Melis2019}
L.~{Melis}, C.~{Song}, E.~{De Cristofaro}, and V.~{Shmatikov}, ``Exploiting unintended feature leakage in collaborative learning,'' in \emph{2019 IEEE Symposium on Security and Privacy (SP)}, 2019, pp. 691--706.

\bibitem{9735364}
C.~Chen, L.~Lyu, H.~Yu, and G.~Chen, ``Practical attribute reconstruction attack against federated learning,'' \emph{IEEE Transactions on Big Data}, vol.~10, no.~6, pp. 851--863, 2024.

\bibitem{10024757}
J.~Geng, Y.~Mou, Q.~Li, F.~Li, O.~Beyan, S.~Decker, and C.~Rong, ``Improved gradient inversion attacks and defenses in federated learning,'' \emph{IEEE Transactions on Big Data}, vol.~10, no.~6, pp. 839--850, 2024.

\bibitem{truex2020ldp}
S.~Truex, L.~Liu, K.-H. Chow, M.~E. Gursoy, and W.~Wei, ``Ldp-fed: Federated learning with local differential privacy,'' in \emph{Proceedings of the third ACM International Workshop on Edge Systems, Analytics and Networking}, 2020, pp. 61--66.

\bibitem{WU2022362}
\BIBentryALTinterwordspacing
X.~Wu, Y.~Zhang, M.~Shi, P.~Li, R.~Li, and N.~N. Xiong, ``An adaptive federated learning scheme with differential privacy preserving,'' \emph{Future Generation Computer Systems}, vol. 127, pp. 362--372, 2022. [Online]. Available: \url{https://www.sciencedirect.com/science/article/pii/S0167739X21003617}
\BIBentrySTDinterwordspacing

\bibitem{mcmahan2018LSTM}
H.~B. McMahan, D.~Ramage, K.~Talwar, and L.~Zhang, ``Learning differentially private recurrent language models,'' \emph{CoRR}, vol. abs/1710.06963, 2018.

\bibitem{Wei2020}
K.~Wei, J.~Li, M.~Ding, C.~Ma, H.~H. Yang, F.~Farokhi, S.~Jin, T.~Q.~S. Quek, and H.~V. Poor, ``Federated learning with differential privacy: Algorithms and performance analysis,'' \emph{IEEE Transactions on Information Forensics and Security}, vol.~15, pp. 3454--3469, 2020.

\bibitem{wu2019value}
N.~Wu, F.~Farokhi, D.~Smith, and M.~A. Kaafar, ``The value of collaboration in convex machine learning with differential privacy,'' \emph{CoRR}, vol. abs/1906.09679, 2019.

\bibitem{9935302}
G.~Xu, X.~Han, S.~Xu, T.~Zhang, H.~Li, X.~Huang, and R.~H. Deng, ``Hercules: Boosting the performance of privacy-preserving federated learning,'' \emph{IEEE Transactions on Dependable and Secure Computing}, vol.~20, no.~5, pp. 4418--4433, 2023.

\bibitem{9833648}
H.~Tian, C.~Zeng, Z.~Ren, D.~Chai, J.~Zhang, K.~Chen, and Q.~Yang, ``Sphinx: Enabling privacy-preserving online learning over the cloud,'' in \emph{2022 IEEE Symposium on Security and Privacy}, 2022, pp. 2487--2501.

\bibitem{spindle}
D.~Froelicher, J.~R. Troncoso-Pastoriza, A.~Pyrgelis, S.~Sav, J.~S. Sousa, J.-P. Bossuat, and J.-P. Hubaux, ``Scalable privacy-preserving distributed learning,'' \emph{PETS}, 2021.

\bibitem{poseidon}
S.~Sav, A.~Pyrgelis, J.~R. Troncoso-Pastoriza, D.~Froelicher, J.-P. Bossuat, J.~S. Sousa, and J.-P. Hubaux, ``Poseidon: Privacy-preserving federated neural network learning,'' in \emph{NDSS}, 2021.

\bibitem{rhode}
S.~Sav, A.~Diaa, A.~Pyrgelis, J.-P. Bossuat, and J.-P. Hubaux, ``Privacy-preserving federated recurrent neural networks,'' \emph{PoPETs}, no.~4, pp. 500--521, 2021.

\bibitem{Bozdemir}
B.~Bozdemir, B.~A. {\"O}zdemir, and M.~{\"O}nen, ``Prida: Privacy-preserving data aggregation with multiple data customers,'' in \emph{ICT Systems Security and Privacy Protection}, N.~Pitropakis, S.~Katsikas, S.~Furnell, and K.~Markantonakis, Eds.\hskip 1em plus 0.5em minus 0.4em\relax Springer Nature Switzerland, 2024, pp. 46--60.

\bibitem{10654543}
X.~Yang, Z.~Liu, X.~Tang, R.~Lu, and B.~Liu, ``An efficient and multi-private key secure aggregation scheme for federated learning,'' \emph{IEEE Transactions on Services Computing}, vol.~17, no.~5, pp. 1998--2011, 2024.

\bibitem{zheng2019helen}
W.~Zheng, R.~A. Popa, J.~E. Gonzalez, and I.~Stoica, ``Helen: Maliciously secure coopetitive learning for linear models,'' in \emph{IEEE S\&P}, 2019.

\bibitem{9187932}
Y.~Li, Y.~Zhou, A.~Jolfaei, D.~Yu, G.~Xu, and X.~Zheng, ``Privacy-preserving federated learning framework based on chained secure multiparty computing,'' \emph{IEEE Internet of Things Journal}, vol.~8, no.~8, pp. 6178--6186, 2021.

\bibitem{falcon}
S.~Wagh, S.~Tople, F.~Benhamouda, E.~Kushilevitz, P.~Mittal, and T.~Rabin, ``{FALCON: H}onest-majority maliciously secure framework for private deep learning,'' \emph{PETS}, 2020.

\bibitem{9139658}
R.~Kanagavelu, Z.~Li, J.~Samsudin, Y.~Yang, F.~Yang, R.~S. Mong~Goh, M.~Cheah, P.~Wiwatphonthana, K.~Akkarajitsakul, and S.~Wang, ``Two-phase multi-party computation enabled privacy-preserving federated learning,'' in \emph{2020 20th IEEE/ACM International Symposium on Cluster, Cloud and Internet Computing (CCGRID)}, 2020, pp. 410--419.

\bibitem{CHEN2024120481}
\BIBentryALTinterwordspacing
L.~Chen, D.~Xiao, Z.~Yu, and M.~Zhang, ``Secure and efficient federated learning via novel multi-party computation and compressed sensing,'' \emph{Information Sciences}, vol. 667, p. 120481, 2024. [Online]. Available: \url{https://www.sciencedirect.com/science/article/pii/S0020025524003943}
\BIBentrySTDinterwordspacing

\bibitem{10646833}
N.~Jawalkar, K.~Gupta, A.~Basu, N.~Chandran, D.~Gupta, and R.~Sharma, ``Orca: Fss-based secure training and inference with gpus,'' in \emph{2024 IEEE Symposium on Security and Privacy}, 2024, pp. 597--616.

\bibitem{9900067}
C.~Zhang, S.~Ekanut, L.~Zhen, and Z.~Li, ``Augmented multi-party computation against gradient leakage in federated learning,'' \emph{IEEE Transactions on Big Data}, vol.~10, no.~6, pp. 742--751, 2024.

\bibitem{securefedyj}
T.~Marchand, B.~Muzellec, C.~Beguier, J.~O.~d. Terrail, and M.~Andreux, ``Securefedyj: a safe feature gaussianization protocol for federated learning,'' in \emph{Proceedings of the 36th International Conference on Neural Information Processing Systems}, ser. NIPS '22.\hskip 1em plus 0.5em minus 0.4em\relax Red Hook, NY, USA: Curran Associates Inc., 2024.

\bibitem{median}
G.~Aggarwal, N.~Mishra, and B.~Pinkas, ``Secure computation of the median (and other elements of specified ranks),'' \emph{J. Cryptology}, vol.~23, pp. 373--401, 2010.

\bibitem{healthcare_survey}
\BIBentryALTinterwordspacing
F.~Zhang, D.~Kreuter, Y.~Chen, S.~Dittmer, S.~Tull, T.~Shadbahr, M.~Schut, F.~Asselbergs, S.~Kar, S.~Sivapalaratnam, S.~Williams, M.~Koh, Y.~Henskens, B.~{de Wit}, U.~D’Alessandro, B.~Bah, O.~Secka, P.~Nachev, R.~Gupta, S.~Trompeter, N.~Boeckx, C.~{van Laer}, G.~A. Awandare, K.~Sarpong, L.~Amenga-Etego, M.~Leers, M.~Huijskens, S.~McDermott, W.~H. Ouwehand, J.~Rudd, C.-B. Schonlieb, N.~Gleadall, M.~Roberts, J.~Preller, J.~H. Rudd, J.~A. Aston, C.-B. Schonlieb, N.~Gleadall, and M.~Roberts, ``Recent methodological advances in federated learning for healthcare,'' \emph{Patterns}, vol.~5, no.~6, p. 101006, 2024. [Online]. Available: \url{https://www.sciencedirect.com/science/article/pii/S2666389924001314}
\BIBentrySTDinterwordspacing

\bibitem{local_example1}
\BIBentryALTinterwordspacing
R.~Kerkouche, G.~\'{A}cs, C.~Castelluccia, and P.~Genev\`{e}s, ``Privacy-preserving and bandwidth-efficient federated learning: an application to in-hospital mortality prediction,'' in \emph{Proceedings of the Conference on Health, Inference, and Learning}, ser. CHIL '21.\hskip 1em plus 0.5em minus 0.4em\relax New York, NY, USA: Association for Computing Machinery, 2021, p. 25–35. [Online]. Available: \url{https://doi.org/10.1145/3450439.3451859}
\BIBentrySTDinterwordspacing

\bibitem{local_example2}
B.~Camajori~Tedeschini, S.~Savazzi, R.~Stoklasa, L.~Barbieri, I.~Stathopoulos, M.~Nicoli, and L.~Serio, ``Decentralized federated learning for healthcare networks: A case study on tumor segmentation,'' \emph{IEEE Access}, vol.~10, pp. 8693--8708, 2022.

\bibitem{local_example3}
S.~Han, H.~Ding, S.~Zhao, S.~Ren, Z.~Wang, J.~Lin, and S.~Zhou, ``Practical and robust federated learning with highly scalable regression training,'' \emph{IEEE Transactions on Neural Networks and Learning Systems}, vol.~35, no.~10, pp. 13\,801--13\,815, 2024.

\bibitem{global_example1}
Z.~Lian, Q.~Yang, W.~Wang, Q.~Zeng, M.~Alazab, H.~Zhao, and C.~Su, ``Deep-fel: Decentralized, efficient and privacy-enhanced federated edge learning for healthcare cyber physical systems,'' \emph{IEEE Transactions on Network Science and Engineering}, vol.~9, no.~5, pp. 3558--3569, 2022.

\bibitem{global_example2}
J.~H. Yoo, H.~M. Son, H.~Jeong, E.-H. Jang, A.~Y. Kim, H.~Y. Yu, H.~J. Jeon, and T.-M. Chung, ``Personalized federated learning with clustering: Non-iid heart rate variability data application,'' in \emph{2021 International Conference on Information and Communication Technology Convergence (ICTC)}, 2021, pp. 1046--1051.

\bibitem{batch1}
\BIBentryALTinterwordspacing
Z.~Du, J.~Sun, A.~Li, P.-Y. Chen, J.~Zhang, H.~H. Li, and Y.~Chen, ``Rethinking normalization methods in federated learning,'' in \emph{Proceedings of the 3rd International Workshop on Distributed Machine Learning}, ser. DistributedML '22.\hskip 1em plus 0.5em minus 0.4em\relax New York, NY, USA: Association for Computing Machinery, 2022, p. 16–22. [Online]. Available: \url{https://doi.org/10.1145/3565010.3569062}
\BIBentrySTDinterwordspacing

\bibitem{wang2023batch}
Y.~Wang, Q.~Shi, and T.-H. Chang, ``Why batch normalization damage federated learning on non-iid data?'' \emph{IEEE Transactions on Neural Networks and Learning Systems}, vol.~36, no.~1, pp. 1692--1706, 2025.

\bibitem{batch3}
B.~Casella, R.~Esposito, A.~Sciarappa, C.~Cavazzoni, and M.~Aldinucci, ``Experimenting with normalization layers in federated learning on non-iid scenarios,'' \emph{CoRR}, vol. abs/2303.10630, 2023.

\bibitem{median_implementation}
\BIBentryALTinterwordspacing
C.~Goelz, S.~Vieluf, and H.~Ballhausen, ``A secure median implementation for the federated secure computing architecture,'' \emph{Applied Sciences}, 2024. [Online]. Available: \url{https://api.semanticscholar.org/CorpusID:272460087}
\BIBentrySTDinterwordspacing

\bibitem{noniid}
\BIBentryALTinterwordspacing
Q.~Li, Y.~Diao, Q.~Chen, and B.~He, ``Federated learning on non-iid data silos: An experimental study,'' \emph{2022 IEEE 38th International Conference on Data Engineering (ICDE)}, pp. 965--978, 2021. [Online]. Available: \url{https://api.semanticscholar.org/CorpusID:231786564}
\BIBentrySTDinterwordspacing

\bibitem{yeojohnson}
\BIBentryALTinterwordspacing
I.-K. Yeo and R.~A. Johnson, ``A new family of power transformations to improve normality or symmetry,'' \emph{Biometrika}, vol.~87, no.~4, pp. 954--959, 2000. [Online]. Available: \url{http://www.jstor.org/stable/2673623}
\BIBentrySTDinterwordspacing

\bibitem{boxcox}
\BIBentryALTinterwordspacing
G.~E.~P. Box and D.~R. Cox, ``An analysis of transformations,'' \emph{Journal of the royal statistical society series b-methodological}, vol.~26, pp. 211--243, 1964. [Online]. Available: \url{https://api.semanticscholar.org/CorpusID:15192218}
\BIBentrySTDinterwordspacing

\bibitem{layernormalization}
J.~L. Ba, J.~R. Kiros, and G.~E. Hinton, ``Layer normalization,'' \emph{CoRR}, vol. abs/1607.06450, 2016.

\bibitem{batchnormalization}
\BIBentryALTinterwordspacing
S.~Ioffe and C.~Szegedy, ``Batch normalization: Accelerating deep network training by reducing internal covariate shift,'' in \emph{Proceedings of the 32nd International Conference on Machine Learning}, ser. Proceedings of Machine Learning Research, F.~Bach and D.~Blei, Eds., vol.~37.\hskip 1em plus 0.5em minus 0.4em\relax Lille, France: PMLR, 07--09 Jul 2015, pp. 448--456. [Online]. Available: \url{https://proceedings.mlr.press/v37/ioffe15.html}
\BIBentrySTDinterwordspacing

\bibitem{instancenormalization}
D.~Ulyanov, A.~Vedaldi, and V.~Lempitsky, ``Instance normalization: The missing ingredient for fast stylization,'' \emph{CoRR}, vol. abs/1607.08022, 2016.

\bibitem{groupnormalization}
Y.~Wu and K.~He, ``Group normalization,'' in \emph{Proceedings of the European Conference on Computer Vision (ECCV)}, 2018, pp. 3--19.

\bibitem{noniidIoT}
Z.~Lu, H.~Pan, Y.~Dai, X.~Si, and Y.~Zhang, ``Federated learning with non-iid data: A survey,'' \emph{IEEE Internet of Things Journal}, vol.~11, no.~11, pp. 19\,188--19\,209, 2024.

\bibitem{10833754}
C.~Chen, T.~Liao, X.~Deng, Z.~Wu, S.~Huang, and Z.~Zheng, ``Advances in robust federated learning: A survey with heterogeneity considerations,'' \emph{IEEE Transactions on Big Data}, vol.~11, no.~3, pp. 1548--1567, 2025.

\bibitem{9826416}
Y.~He, Y.~Chen, X.~Yang, H.~Yu, Y.-H. Huang, and Y.~Gu, ``Learning critically: Selective self-distillation in federated learning on non-iid data,'' \emph{IEEE Transactions on Big Data}, vol.~10, no.~6, pp. 789--800, 2024.

\bibitem{fully_homomorphic_encryption}
\BIBentryALTinterwordspacing
C.~Gentry, ``Fully homomorphic encryption using ideal lattices,'' in \emph{Proceedings of the Forty-First Annual ACM Symposium on Theory of Computing}, ser. STOC '09.\hskip 1em plus 0.5em minus 0.4em\relax New York, NY, USA: Association for Computing Machinery, 2009, p. 169–178. [Online]. Available: \url{https://doi.org/10.1145/1536414.1536440}
\BIBentrySTDinterwordspacing

\bibitem{cheon2017homomorphic}
J.~H. Cheon, A.~Kim, M.~Kim, and Y.~Song, ``Homomorphic encryption for arithmetic of approximate numbers,'' in \emph{ASIACRYPT}, 2017.

\bibitem{mouchet2021multiparty}
C.~Mouchet, J.~Troncoso-Pastoriza, J.-P. Bossuat, and J.-P. Hubaux, ``Multiparty homomorphic encryption from ring-learning-with-errors,'' \emph{Proceedings on Privacy Enhancing Technologies}, vol. 2021, no.~4, pp. 291--311, 2021.

\bibitem{Bossuat2021}
J.-P. Bossuat, C.~Mouchet, J.~Troncoso-Pastoriza, and J.-P. Hubaux, ``Efficient bootstrapping for approximate homomorphic encryption with non-sparse keys,'' in \emph{EUROCRYPT}, 2021, pp. 587--617.

\bibitem{flower}
D.~J. Beutel, T.~Topal, A.~Mathur, X.~Qiu, J.~Fernandez-Marques, Y.~Gao, L.~Sani, H.~L. Kwing, T.~Parcollet, P.~P.~d. Gusmão, and N.~D. Lane, ``Flower: A friendly federated learning research framework,'' \emph{CoRR}, vol. abs/2007.14390, 2020.

\bibitem{cifarPaper}
A.~Krizhevsky, ``Learning multiple layers of features from tiny images,'' \emph{Technical Report, University of Toronto}, 2012.

\bibitem{MNIST}
Y.~LeCun and C.~Cortes, ``{MNIST} handwritten digit database,'' 2010.

\bibitem{bcw}
\BIBentryALTinterwordspacing
W.~H. Wolberg, W.~N. Street, and O.~L. Mangasarian, ``Breast cancer wisconsin (diagnostic) data set,'' 1995. [Online]. Available: \url{https://archive.ics.uci.edu/dataset/17/breast+cancer+wisconsin+diagnostic}
\BIBentrySTDinterwordspacing

\bibitem{hepatisis}
\BIBentryALTinterwordspacing
G.~Ayana, P.~Smets, S.~Destercke, and J.~Ma, ``Hcv data set,'' 2019. [Online]. Available: \url{https://archive.ics.uci.edu/dataset/571/hcv+data}
\BIBentrySTDinterwordspacing

\bibitem{parkinsons}
\BIBentryALTinterwordspacing
A.~Tsanas, M.~A. Little, P.~E. McSharry, J.~Spielman, and L.~O. Ramig, ``Parkinson's telemonitoring data set,'' 2009. [Online]. Available: \url{https://archive.ics.uci.edu/dataset/189/parkinsons+telemonitoring}
\BIBentrySTDinterwordspacing

\bibitem{fhebnorm}
A.~Ibarrondo and M.~{\"O}nen, ``Fhe-compatible batch normalization for privacy preserving deep learning,'' in \emph{Data Privacy Management, Cryptocurrencies and Blockchain Technology: ESORICS 2018 International Workshops, DPM 2018 and CBT 2018, Barcelona, Spain, September 6-7, 2018, Proceedings 13}.\hskip 1em plus 0.5em minus 0.4em\relax Springer, 2018, pp. 389--404.

\bibitem{lattigo}
``Lattigo v6,'' Online: \url{https://github.com/tuneinsight/lattigo}, Aug. 2024, {E}PFL-LDS, Tune Insight SA.

\bibitem{homomorphicInverse}
\BIBentryALTinterwordspacing
J.~H. Cheon, W.~Kim, and J.~H. Park, ``Efficient homomorphic evaluation on large intervals,'' Cryptology {ePrint} Archive, Paper 2022/280, 2022. [Online]. Available: \url{https://eprint.iacr.org/2022/280}
\BIBentrySTDinterwordspacing

\bibitem{goldschmidt}
R.~Goldschmidt, ``Applications of division by convergence,'' 08 2005.

\bibitem{openai2024gpt4o}
\BIBentryALTinterwordspacing
OpenAI, ``Gpt-4o: Multimodal ai model,'' 2024, accessed: February 28, 2025. [Online]. Available: \url{https://openai.com/research/gpt-4o}
\BIBentrySTDinterwordspacing

\end{thebibliography}

\vfill

\clearpage








\section*{SUPPLEMENTARY MATERIAL}

\subsection{Detailed Model and Dataset Information}\label{sec:supdetailedModel}

Table~\ref{tab:model_architectures2} summarizes the model architectures, while Table~\ref{tab:dataset_summary} provides an overview of the datasets used in the experiments presented in Section IV-B of the main manuscript.

\begin{table*}[t]
    \centering
    \caption{Detailed Model Architectures for Each Dataset. For CIFAR10 with 50 clients, a reduced model was used due to hardware limitations.}
    \label{tab:model_architectures2}
    \includegraphics[width=0.9\textwidth]{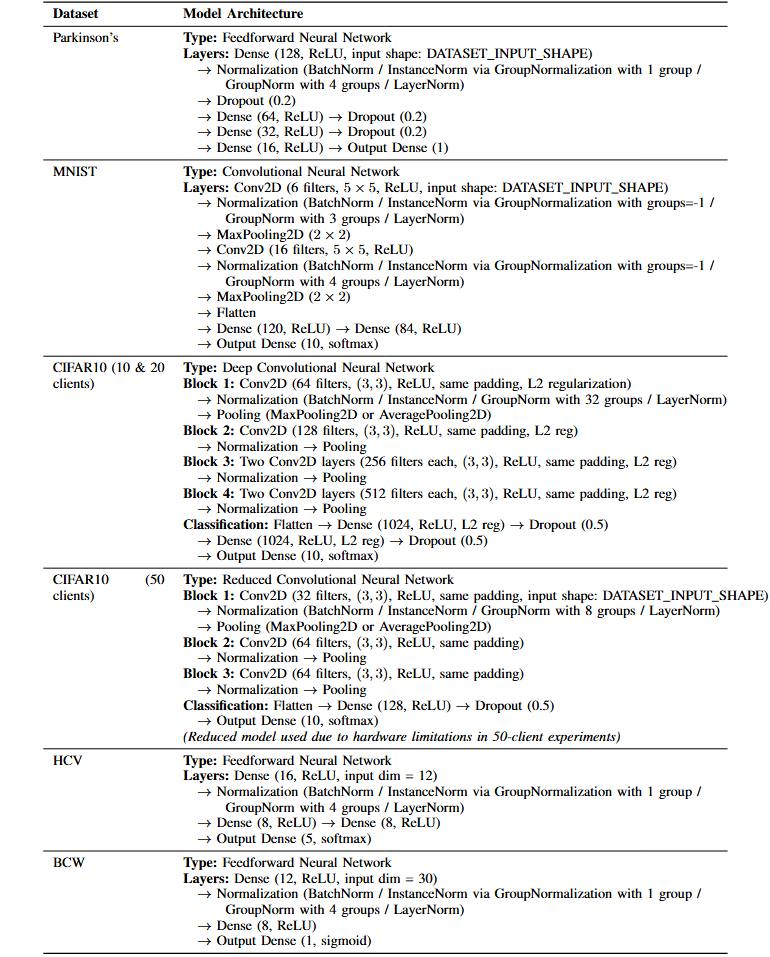}
\end{table*}

\subsection{Details of PPF MinMax Scaling}\label{sec:supminmax}

This protocol begins by key generation (Line 1 of Algorithm~\ref{alg:federated_minmax}). Each party then computes local feature-wise minima and maxima ($X_{\min,p}$, $X_{\max,p}$) across their dataset (Lines 3–7), producing two arrays of length $N$, where $N$ is the number of features. These values are encrypted using $pk$ to obtain ciphertexts $\langle X_{\min,p} \rangle$ and $\langle X_{\max,p} \rangle$, where each feature's value is stored in a separate slot of a ciphertext (Lines 8--9). The encrypted values are then sent to the server (Line 10).

\begin{algorithm}[H]
\caption{PPF MinMax Scaling Protocol (\textsf{PPFMinMax()})}
\label{alg:federated_minmax}
\small
\begin{algorithmic}[1]

\STATE \textbf{Initialization:}  $\{sk_p\}, pk \gets \textsf{KeyGen($\lambda$)}$

\STATE \textbf{Compute Local MinMax (Client Side):}
\FOR{each party $p$ \textbf{in} $\{1, \ldots, P\}$}
    \FOR{each feature $j$ \textbf{in} $\{1, \ldots, N\}$}
        \STATE $X_{\min,p}[j] \gets \min(D_p[j])$
        \STATE $X_{\max,p}[j] \gets \max(D_p[j])$
    \ENDFOR
    \STATE $\langle X_{\min,p} \rangle$ $\gets$ \textsf{Encrypt($X_{\min,p}$, $pk$)}
    \STATE $\langle X_{\max,p} \rangle$ $\gets$ \textsf{Encrypt($X_{\max,p}$, $pk$)}
    \STATE Send $\langle X_{\min,p} \rangle$ and $\langle X_{\max,p} \rangle$ to the server
\ENDFOR

\STATE \textbf{Normalization Factor Selection:}
\STATE Parties agree on a normalization vector $V_{abs}$

\STATE \textbf{Global MinMax Computation (Server Side):}
\STATE Normalize received ciphertexts: $\langle X_{\min,p} \rangle \gets \langle X_{\min,p} \rangle \odot V_{abs}$, $\langle X_{\max,p} \rangle \gets \langle X_{\max,p} \rangle \odot V_{abs}$
\STATE Initialize $\langle X_{\min,g} \rangle$ and $\langle X_{\max,g} \rangle$ using the normalized values of the first party
\FOR{each remaining party $p$ \textbf{in} $\{2, \ldots, P\}$}
    \STATE $\langle X_{\min,g} \rangle \gets \textsf{Min(}\langle X_{\min,g} \rangle, \langle X_{\min,p} \rangle \textsf{)}$
    \STATE $\langle X_{\max,g} \rangle \gets \textsf{Max(}\langle X_{\max,g} \rangle, \langle X_{\max,p} \rangle\textsf{)}$
    \STATE $\langle X_{\min,g} \rangle \gets \textsf{CBootstrap(}\langle X_{\min,g} \rangle, \{sk_p\}\textsf{)}$
    \STATE $\langle X_{\max,g} \rangle \gets \textsf{CBootstrap(}\langle X_{\max,g} \rangle, \{sk_p\}\textsf{)}$
\ENDFOR
\STATE Restore original scale: $\langle X_{\min,g} \rangle \gets \langle X_{\min,g} \rangle \odot V_{abs}^{-1}$, $\langle X_{\max,g} \rangle \gets \langle X_{\max,g} \rangle \odot V_{abs}^{-1}$

\STATE \textbf{Decryption and Local Normalization (Client Side):}
\STATE $X_{\min,g}$ $\gets$ \textsf{CDecrypt($\langle X_{\min,g} \rangle$, \{$sk_p$\})}
\STATE $X_{\max,g}$ $\gets$ \textsf{CDecrypt($\langle X_{\max,g} \rangle$, \{$sk_p$\})}
\STATE Parties apply MinMax scaling locally using $X_{\min,g}$,$X_{\max,g}$

\end{algorithmic}
\end{algorithm}

Before computation, the parties collaboratively agree on a normalization vector $V_{abs}$, which contains estimated feature-wise absolute maximum values. (Line 13). This vector scales the minimum and maximum values into the range $[-1,1]$. Assuming an absolute maximum is reasonable for most ML datasets, since feature ranges—such as 'age' or 'weight'—are typically well-defined. This value serves only for temporary scaling during normalization and does not need to be exact, as the original scale is restored afterward.

Upon receiving the encrypted minimum and maximum values from all parties, the server scales them using $V_{abs}$ to ensure they fall within the required range by utilizing homomorphic multiplication (Line 15). The server initializes global minimum ($\langle X_{\min,g} \rangle$) and global maximum ($\langle X_{\max,g} \rangle$) ciphertexts using the normalized values from the first party (Line 16). It then iteratively compares $\langle X_{\min,g} \rangle$ with each party's encrypted minimum values using homomorphic comparison and updates $\langle X_{\min,g} \rangle$. After each comparison, collective bootstrapping is used to refresh ciphertext levels. The same process is used to compute $\langle X_{\max,g} \rangle$ through homomorphic maximum comparison (Lines 17--22). Once the $\langle X_{\min,g} \rangle$ and $\langle X_{\max,g} \rangle$ values are determined, the server rescales them by multiplying with the inverse of the normalization vector, $V_{abs}^{-1}$, to restore the original range (Line 23). The final encrypted results are sent back to the parties, who decrypt them collaboratively to obtain $X_{\min,g}$ and $X_{\max,g}$ (Lines 25--26). Each party then applies MinMax scaling to its local dataset (Line 27).

\subsection{Supplementary Figures and Tables}\label{sec:supFigures}

In this section, we provide the supplementary figures and tables for our experiments. 

\begin{figure*}[h]
\centering
\includegraphics[width=0.9\textwidth]{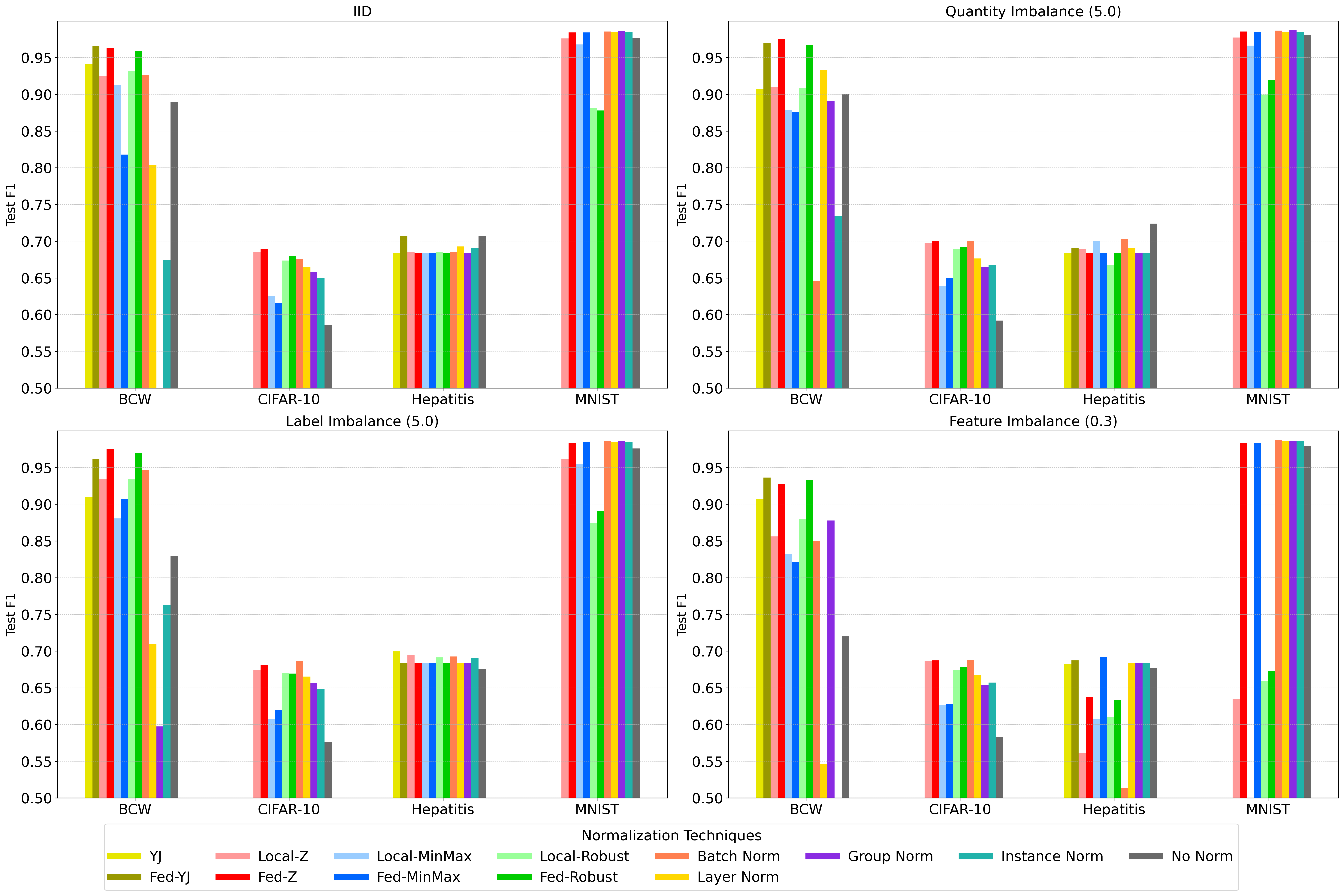}
\caption{Test F1 scores of different imbalance settings for 30 clients (50 for image datasets). Less imbalanced (less heterogeneous) parameters are selected for each imbalance type. Only classification datasets are included due to representation with F1 score metric.}
\label{fig:grouped_plot_low_skew}
\end{figure*}

\begin{figure}[h]
\centering
\includegraphics[width=0.8\columnwidth]{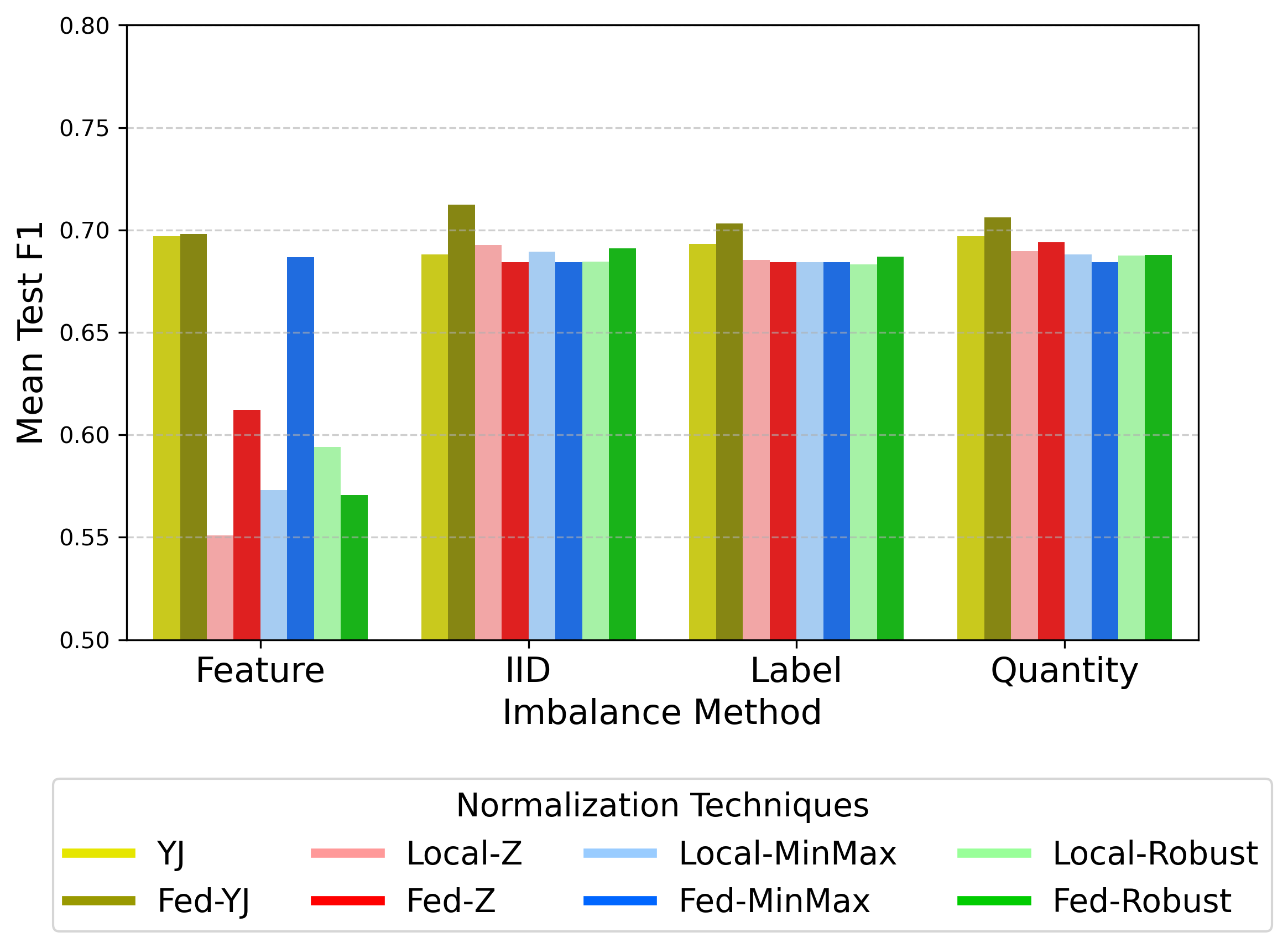}
\caption{Test F1 results of Hepatitis dataset averaged over client numbers to observe the impact of different heterogeneous scenarios.}
\label{fig:hepatitis_skew}
\end{figure}

\begin{figure}[h]
\centering
\includegraphics[width=0.8\columnwidth]{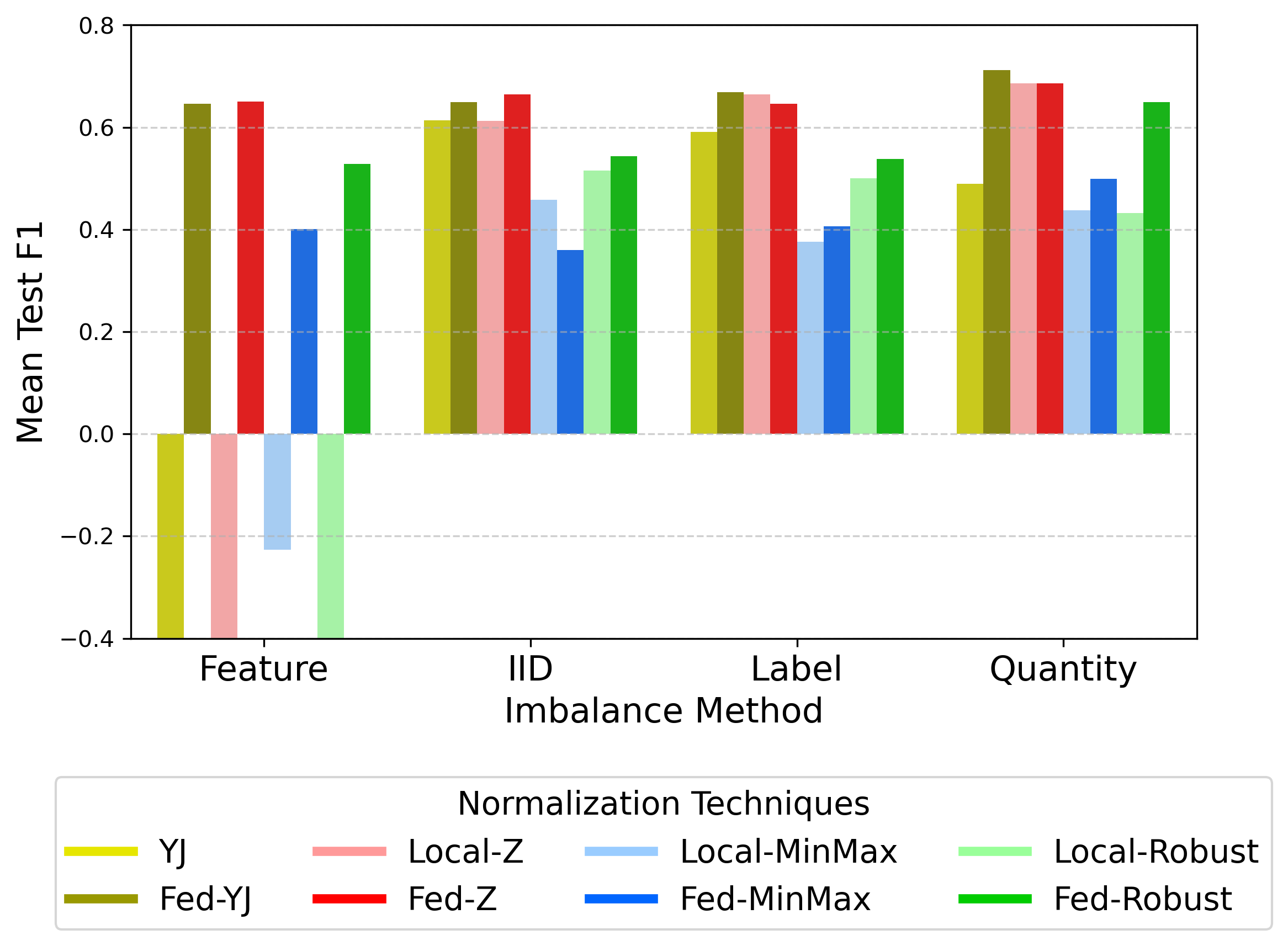}
\caption{Test $R^2$ results of Parkinson's dataset averaged over client numbers to observe the impact of different heterogeneous scenarios.}
\label{fig:parkinson_skew}
\end{figure}

\begin{figure}[h]
\centering
\includegraphics[width=0.8\columnwidth]{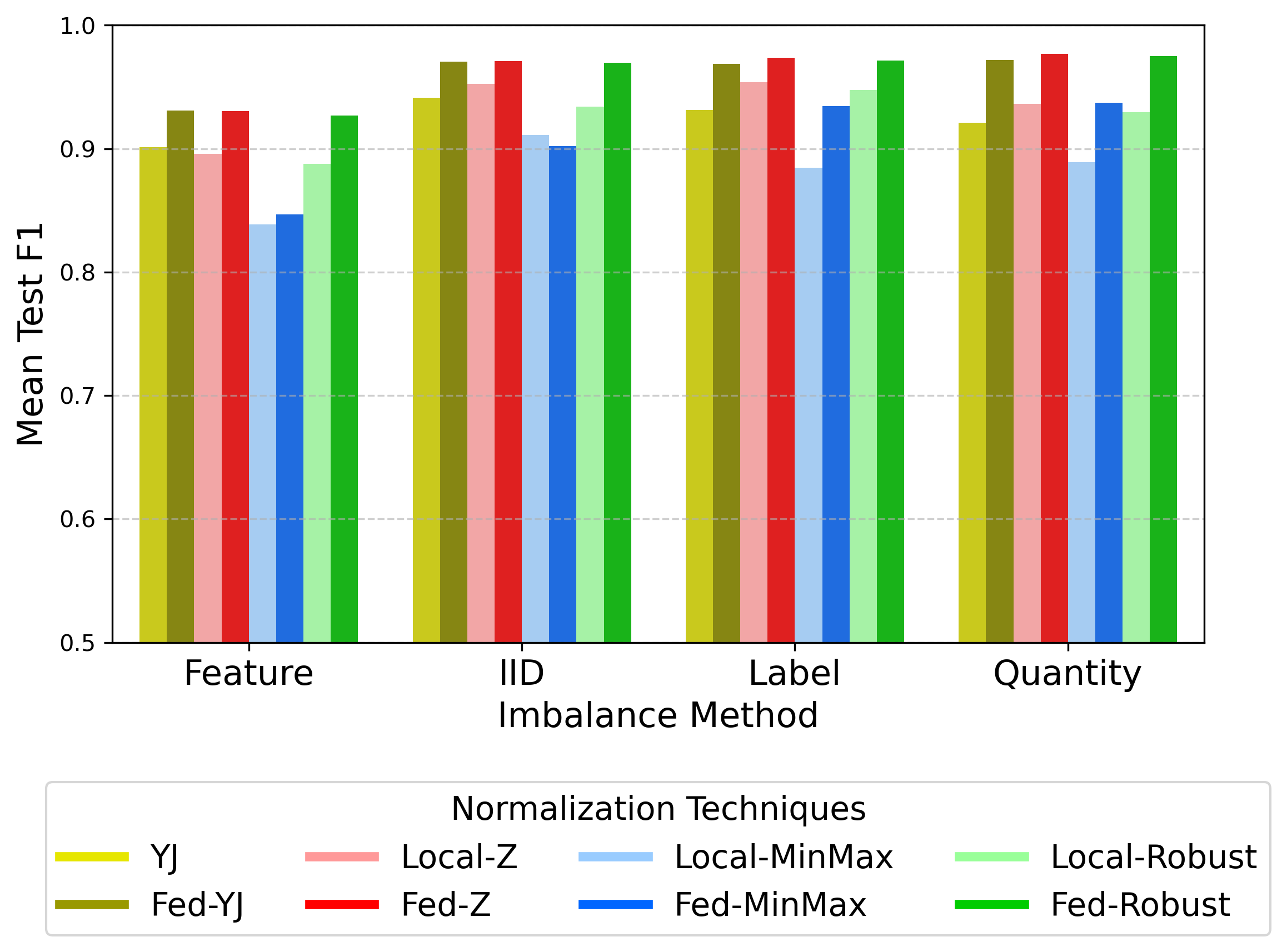}
\caption{Test F1 results of BCW dataset averaged over client numbers to observe the impact of different heterogeneous scenarios.}
\label{fig:bcw_skew}
\end{figure}

\begin{figure}[h]
\centering
\includegraphics[width=0.75\columnwidth]
{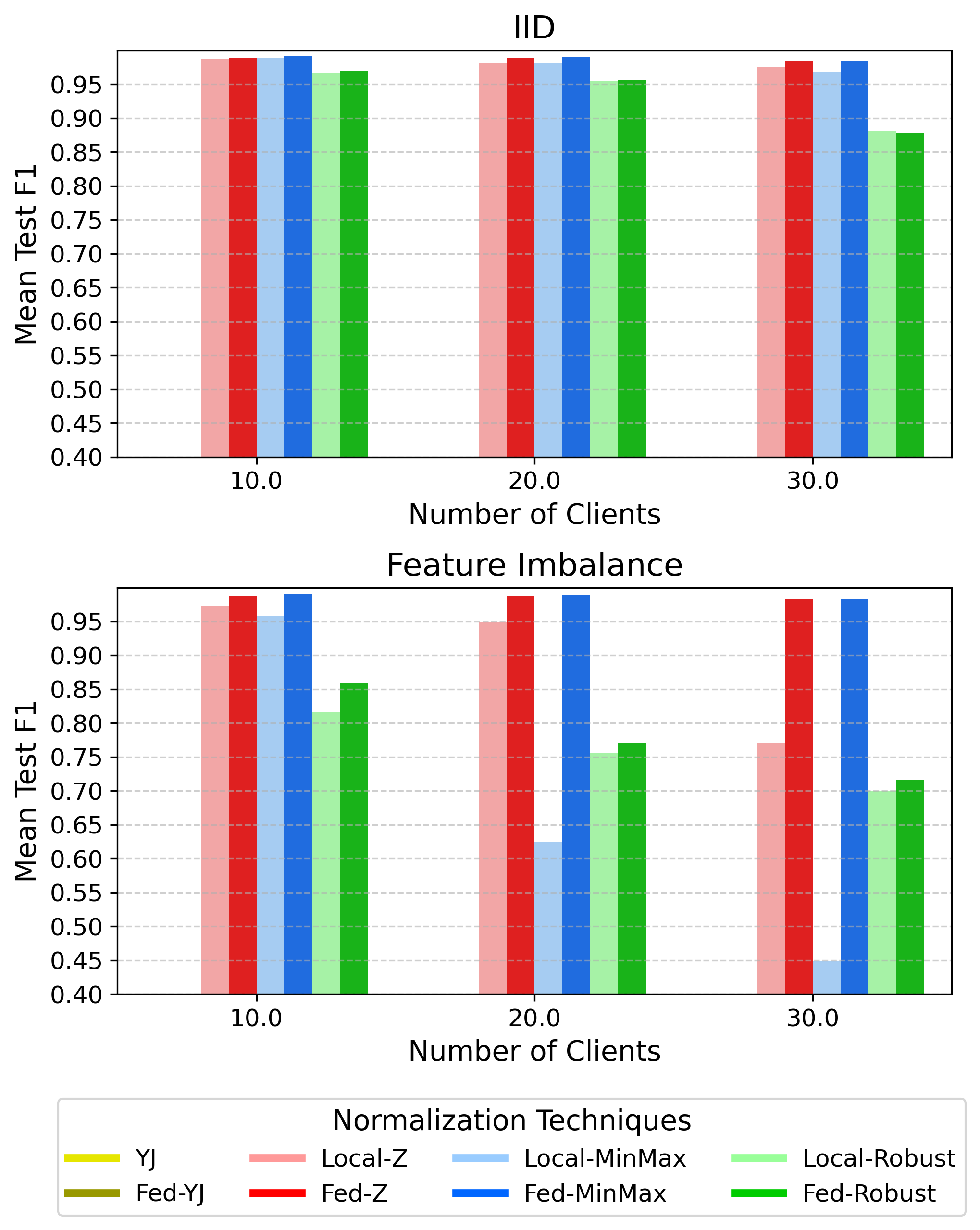}
\caption{Test F1 scores for MNIST dataset for feature imbalance \& IID comparison for observing the impact of number of clients.}
\label{fig:mnist_client}
\end{figure}

\begin{figure}[h]
\centering
\includegraphics[width=0.75\columnwidth]
{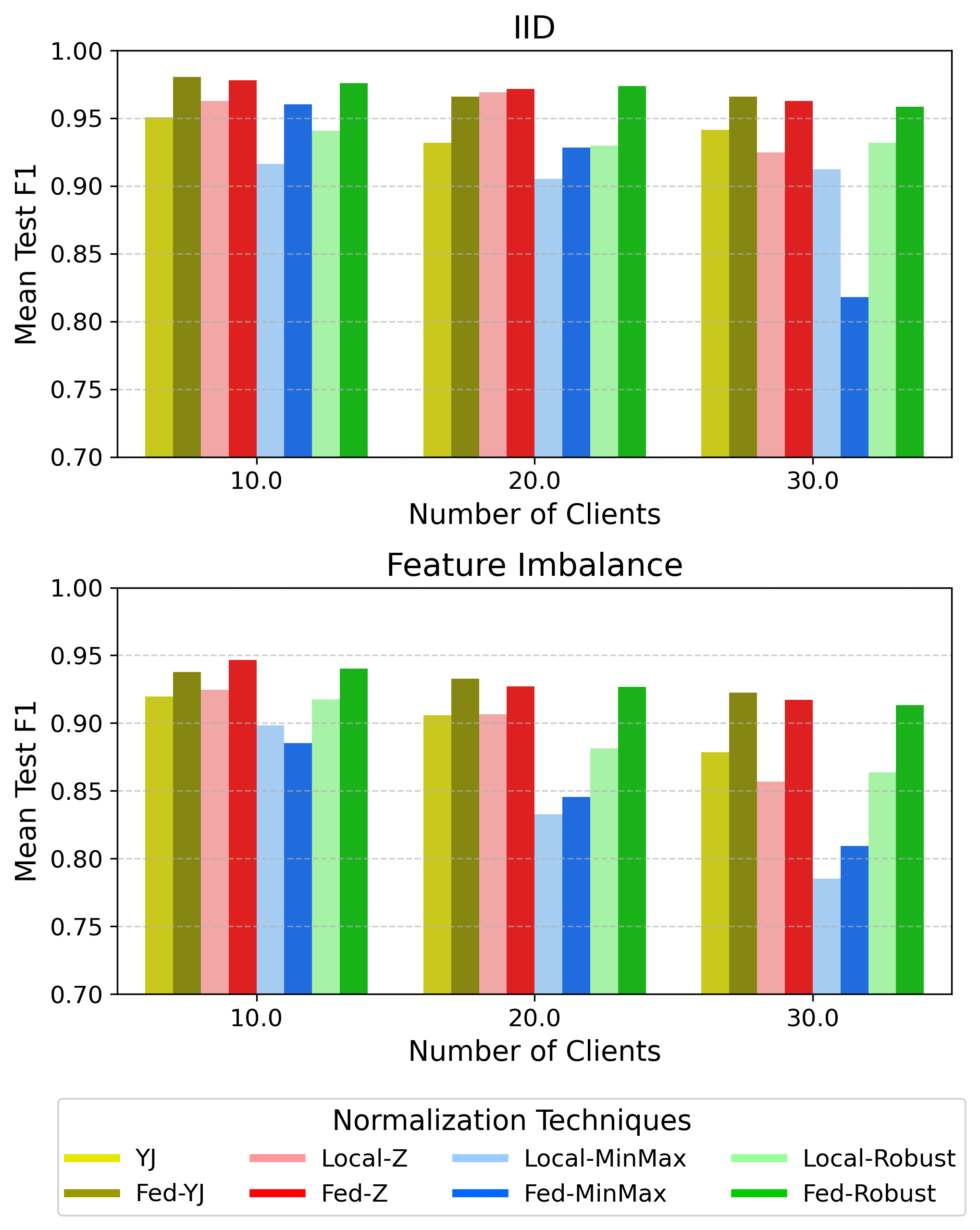}
\caption{Test F1 scores for BCW dataset for feature imbalance \& IID comparison for observing the impact of number of clients.}
\label{fig:bcw_client}
\end{figure}

Figure~\ref{fig:grouped_plot_low_skew} presents the test F1 scores of federated and local normalizations for less heterogeneous scenarios on 30 clients (50 clients for image datasets). 
Figure~\ref{fig:hepatitis_skew},~\ref{fig:parkinson_skew}, and~\ref{fig:bcw_skew}  present the performance of federated and local normalizations for Hepatitis, Parkinson's, and BCW datasets under different heterogeneous FL scenarios, respectively. Figure~\ref{fig:mnist_client} and Figure~\ref{fig:bcw_client} present the impact of the number of clients on federated and local normalizations for MNIST, and BCW datasets. These findings are discussed in Section IV-C in the main manuscript.
\begin{table}[H]
    \centering
    \begin{threeparttable}
    \renewcommand{\arraystretch}{1.2}
    \setlength{\tabcolsep}{6pt}
    
    \caption{Dataset Summary: Type, Sample Count (Features), and Task (Classes)}
    \label{tab:dataset_summary}
    \vspace{1mm}
    \begin{tabular*}{\linewidth}{@{\extracolsep{\fill}}lccc}
        \toprule
        \textbf{Dataset} & \textbf{Type} & \textbf{Count} & \textbf{Task} \\
        \midrule
        HCV           & Tabular & 615 (13)     & Multi-class (5) \\
        Parkinson's   & Tabular & 5,875 (16)   & Regression \\
        BCW           & Tabular & 569 (30)     & Binary \\
        CIFAR-10      & Image   & 60,000 (3,072) & Multi-class (10) \\
        MNIST         & Image   & 70,000 (784) & Multi-class (10) \\
        \bottomrule
    \end{tabular*}
    \end{threeparttable}
\end{table}

\begin{figure}[h]
\centering
\includegraphics[width=\columnwidth]{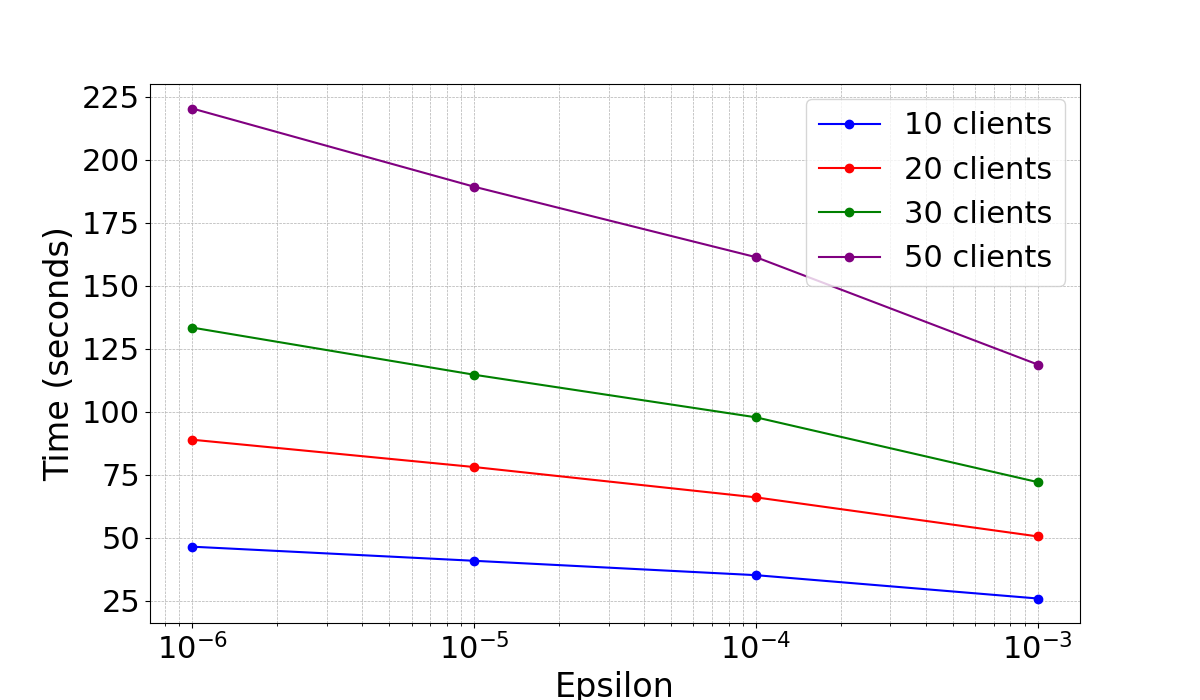}
\caption{Effect of the precision ($\epsilon$) on the PPF $k$-th element algorithm. We observe that runtime increases linearly, while $\epsilon$ decreases exponentially.}
\label{fig:epsilon_impact}
\end{figure}

Additionally, 
Tables~\ref{tab:micro_20clients2},~\ref{tab:micro_30clients2}, and~\ref{tab:micro_50clients2} present the PPF microbenchmarks with 20, 30, and 50 client settings, respectively. Figure~\ref{fig:epsilon_impact} illustrates the empirical impact of the precision value ($\epsilon$) on the PPF $k$-th element algorithm. These tables and figures facilitate the findings and discussion in Section VI-B of the main manuscript.

\begin{table}[t]
    \centering
    \caption{Microbenchmarks with 20 clients}
    \label{tab:micro_20clients2}
    \renewcommand{\arraystretch}{1.2}
    \begin{tabular}{@{}lc@{}}
        \toprule
        \textbf{Operation Name} & \textbf{Time (ms)} \\
        \midrule
        \multicolumn{2}{c}{\textbf{Common Operations}} \\
        \midrule
        Secret Key Generation ($T_{skgen}$) & 9.67 \\
        Collective Public Key Generation ($T_{pkgen}$) & 233.04 \\
        Collective Relinearization Key Generation ($T_{relkgen}$) & 10052.85 \\
        Encryption ($T_{encrypt}$) & 59.28 \\
        Collective Decryption ($T_{decrypt}$) & 949.49 \\
        \cmidrule(lr){1-2}
        \multicolumn{2}{c}{\textbf{PPF Z-Score Normalization}} \\
        \cmidrule(lr){1-2}
        Homomorphic Multiplication ($T_{mul}$) & 173.77 \\
        Homomorphic Summation ($T_{sum}$) & 1.73 \\
        Homomorphic Inverse ($T_{inv}$) & 43269.14 \\
        Collective Bootstrapping ($T_{bootstrap}$) & 1423.71 \\ 
        \cmidrule(lr){1-2}
        \multicolumn{2}{c}{\textbf{PPF MinMax Scaling}} \\
        \cmidrule(lr){1-2}
        Collective Galois Keys generation ($T_{gkgen}$) & 3407.75 \\
        Homomorphic Multiplication ($T_{mul}$) & 173.77 \\
        Homomorphic Inverse ($T_{inv}$) & 43269.14 \\
        Homomorphic Minimum Comparison ($T_{min}$) & 15377.25 \\
        Homomorphic Maximum Comparison ($T_{max}$) & 15486.16 \\
        Collective Bootstrapping ($T_{bootstrap}$) & 1423.71 \\ 
        \cmidrule(lr){1-2}
        \multicolumn{2}{c}{\textbf{PPF Robust Scaling}} \\
        \cmidrule(lr){1-2}
        Homomorphic Summation ($T_{sum}$) & 1.73 \\
        \bottomrule
    \end{tabular}
\end{table}

\begin{table}[t]
    \centering
    \caption{Microbenchmarks with 30 clients}
    \label{tab:micro_30clients2}
    \renewcommand{\arraystretch}{1.2}
    \begin{tabular}{@{}lc@{}}
        \toprule
        \textbf{Operation Name} & \textbf{Time (ms)} \\
        \midrule
        \multicolumn{2}{c}{\textbf{Common Operations}} \\
        \midrule
        Secret Key Generation ($T_{skgen}$) & 9.67 \\
        Collective Public Key Generation ($T_{pkgen}$) & 342.94 \\
        Collective Relinearization Key Generation ($T_{relkgen}$) & 13703.95 \\
        Encryption ($T_{encrypt}$) & 59.28 \\
        Collective Decryption ($T_{decrypt}$) & 1357.74 \\
        \cmidrule(lr){1-2}
        \multicolumn{2}{c}{\textbf{PPF Z-Score Normalization}} \\
        \cmidrule(lr){1-2}
        Homomorphic Multiplication ($T_{mul}$) & 173.77 \\
        Homomorphic Summation ($T_{sum}$) & 1.73 \\
        Homomorphic Inverse ($T_{inv}$) & 56877.70 \\
        Collective Bootstrapping ($T_{bootstrap}$) & 2051.37 \\ 
        \cmidrule(lr){1-2}
        \multicolumn{2}{c}{\textbf{PPF MinMax Scaling}} \\
        \cmidrule(lr){1-2}
        Collective Galois Keys generation ($T_{gkgen}$) & 5244.26 \\
        Homomorphic Multiplication ($T_{mul}$) & 173.77 \\
        Homomorphic Inverse ($T_{inv}$) & 56877.70 \\
        Homomorphic Minimum Comparison ($T_{min}$) & 18146.61 \\
        Homomorphic Maximum Comparison ($T_{max}$) & 18914.46 \\
        Collective Bootstrapping ($T_{bootstrap}$) & 2051.37 \\ 
        \cmidrule(lr){1-2}
        \multicolumn{2}{c}{\textbf{PPF Robust Scaling}} \\
        \cmidrule(lr){1-2}
        Homomorphic Summation ($T_{sum}$) & 1.73 \\
        \bottomrule
    \end{tabular}
\end{table}

\begin{table}[t]
    \centering
    \caption{Microbenchmarks with 50 clients}
    \label{tab:micro_50clients2}
    \renewcommand{\arraystretch}{1.2}
    \begin{tabular}{@{}lc@{}}
        \toprule
        \textbf{Operation Name} & \textbf{Time (ms)} \\
        \midrule
        \multicolumn{2}{c}{\textbf{Common Operations}} \\
        \midrule
        Secret Key Generation ($T_{skgen}$) & 9.67 \\
        Collective Public Key Generation ($T_{pkgen}$) & 567.72 \\
        Collective Relinearization Key Generation ($T_{relkgen}$) & 22486.37 \\
        Encryption ($T_{encrypt}$) & 59.28 \\
        Collective Decryption ($T_{decrypt}$) & 2224.26 \\
        \cmidrule(lr){1-2}
        \multicolumn{2}{c}{\textbf{PPF Z-Score Normalization}} \\
        \cmidrule(lr){1-2}
        Homomorphic Multiplication ($T_{mul}$) & 173.77 \\
        Homomorphic Summation ($T_{sum}$) & 1.73 \\
        Homomorphic Inverse ($T_{inv}$) & 86618.77 \\
        Collective Bootstrapping ($T_{bootstrap}$) & 3263.24 \\
        \cmidrule(lr){1-2}
        \multicolumn{2}{c}{\textbf{PPF MinMax Scaling}} \\
        \cmidrule(lr){1-2}
        Collective Galois Keys generation ($T_{gkgen}$) & 8133.07 \\
        Homomorphic Multiplication ($T_{mul}$) & 173.77 \\
        Homomorphic Inverse ($T_{inv}$) & 86618.77 \\
        Homomorphic Minimum Comparison ($T_{min}$) & 23936.17 \\
        Homomorphic Maximum Comparison ($T_{max}$) & 22475.76 \\
        Collective Bootstrapping ($T_{bootstrap}$) & 3263.24 \\
        \cmidrule(lr){1-2}
        \multicolumn{2}{c}{\textbf{PPF Robust Scaling}} \\
        \cmidrule(lr){1-2}
        Homomorphic Summation ($T_{sum}$) & 1.73 \\
        \bottomrule
    \end{tabular}
\end{table}

\end{document}